\documentclass[notitlepage,11pt,superscriptaddress,amsmath,amssymb,floatfix,reprint,nofootinbib]{revtex4-1}
\usepackage[utf8]{inputenc}
\usepackage{graphicx}
\usepackage[dvipsnames]{xcolor}
\usepackage[colorlinks=true,citecolor=ForestGreen,linkcolor=Red,urlcolor=Blue,hypertexnames=false]{hyperref}
\usepackage[sort&compress]{natbib}
\newcommand{\ie}{\emph{i.e.}}

\newcommand{\eref}[1]{Eq.~(\ref{#1})}
\newcommand{\ee}{\mathrm{e}}
\newcommand{\dd}{\mathrm{d}}
\newcommand{\cmin}{\hat{c}_{\text{min}}}
\newcommand{\fmin}{f_{\text{min}}}
\newcommand{\abs}[1]{\lvert #1\rvert}
\begin{document}
\title{Simple regularities in the dynamics of online news impact}
\author{Matúš Medo}
\affiliation{Institute of Fundamental and Frontier Sciences, University of Electronic Science and Technology of China, Chengdu 610054, PR China}
\affiliation{Department of Radiation Oncology, Inselspital, University Hospital of Bern, and University of Bern, 3010 Bern, Switzerland}
\affiliation{Department of Physics, University of Fribourg, 1700 Fribourg, Switzerland}
\author{Manuel S. Mariani}
\affiliation{Institute of Fundamental and Frontier Sciences, University of Electronic Science and Technology of China, Chengdu 610054, PR China}
\affiliation{URPP Social Networks, Universit\"at Z\"urich, Switzerland}
\author{Linyuan L\"u}
\affiliation{Institute of Fundamental and Frontier Sciences, University of Electronic Science and Technology of China, Chengdu 610054, PR China}
\affiliation{Complex Systems Lab, Beijing Computational Science Research Center, 100193 Beijing, PR China}
\affiliation{Yangtze Delta Region Institute (Huzhou), University of Electronic Science and Technology of China, 313001 Huzhou, PR China}

\begin{abstract}
Online news can quickly reach and affect millions of people, yet we do not know yet whether there exist potential dynamical regularities that govern their impact on the public. We use data from two major news outlets, BBC and New York Times, where the number of user comments can be used as a proxy of news impact. We find that the impact dynamics of online news articles does not exhibit popularity patterns found in many other social and information systems. In particular, we find that a simple exponential distribution yields a better fit to the empirical news impact distributions than a power-law distribution. This observation is explained by the lack or limited influence of the otherwise omnipresent rich-get-richer mechanism in the analyzed data. The temporal dynamics of the news impact exhibits a universal exponential decay which allows us to collapse individual news trajectories into an elementary single curve. We also show how daily variations of user activity directly influence the dynamics of the article impact. Our findings challenge the universal applicability of popularity dynamics patterns found in other social contexts.
\end{abstract}

\maketitle

\section*{Introduction}
Consider a major news, like the results of the presidential elections or the onset of a global epidemic outbreak. In the 80s, we would have discovered it through traditional print and broadcast media. Today, new media and online platforms have disrupted not only the way we discover and consume information, but also the way we form our opinions and attitudes about critical topics for our society like politics~\cite{epstein2015search,aral2019protecting}, science~\cite{brossard2013new,iyengar2019scientific}, and public health~\cite{kata2012anti,johnson2020online}. Online newspapers and social media platforms are now the majour sources of information about events in the world~\cite{stocking2017digital} and provide us with rich data for the study of human attention~\cite{lazer2020studying,yang2020exposure}. Despite the rise of social media, traditional newspapers and mainstream media are still important information sources with large audiences. The importance of mainstrean news sources can be illustrated by, for example, Facebook temporarily increasing their weight in the internal news ranking system in an attempt to respond to misinformation spreading following the U.\ S.\ presidential election.\footnote{\protect\url{https://www.theverge.com/2020/11/24/21612728/facebook-news-feed-us-election-change-mainstream-news-misinformation}}

Most online newspapers allow users to directly comment on news articles~\cite{ksiazek2018commenting}, creating a ``digital public sphere'' where participation is free, recent events are publicly discussed, and comments are visible to everyone~\cite{schafer2015digital}. In such a complex information ecosystem, some news articles impact thousands of users who actively discuss and share them in online platforms\footnote{\protect\url{https://www.nytimes.com/2017/12/20/insider/our-most-commented-on-articles-of-2017.html}}, whereas many others remain little noticed. Therefore, understanding the dynamics of the impact of online news articles is vital not only because it deepens our understanding of how information spreads throughout modern societies, but also because it can potentially help to counteract negative side effects of new media like the spreading of misinformation~\cite{del2016spreading,vosoughi2018spread} and the amplification of ideological segregation~\cite{flaxman2016filter}.

The unprecedented availability of big data on human online activity has allowed us to uncover and model patterns of human behavior and cultural products' popularity in diverse contexts~\cite{lazer2009computational,hofman2017prediction}, revealing universal regularities in the dynamics of cultural products as diverse as scientific papers~\cite{wang2013quantifying}, books~\cite{yucesoy2018success}, and songs~\cite{candia2019universal}, among others. As for online news articles, previous research has unveiled factors that make an online news article more likely to become popular, including story topic~\cite{canter2013misconception}, content emotion~\cite{berger2012makes}, perceived objectivity~\cite{ksiazek2016user}, and format~\cite{liu2015understanding}. Yet, we do not know yet whether there exist universal regularities that govern the dynamics of online news articles' impact. Does the impact of online news articles follow similar patterns as the impact of other types of information items? Is article impact broadly distributed? Are there universal impact patterns for online news articles? How predictable is the dynamics of attention decay for online news articles? To address these questions, we analyze a novel dataset that contains commenting sections of 3,087 articles from the British Broadcasting Corporation (BBC) and a dataset that contains commenting sections of 2,801 articles from the New York Times (NYT).

Previous works have generated insights that generalize well across domains: popularity and impact typically follow heavy-tailed distributions, leading to the emergence of a small number of ``hits''~\cite{thompson2018hit,barabasi2018formula} with disproportionate popularity. These successful outliers emerge from a combination of quality (often referred to as fitness) and social amplification mechanisms such as the rich-get-richer phenomenon~\cite{medo2011temporal}. These regularities in popularity dynamics have been found to govern the popularity and impact dynamics of cultural items as diverse as scientific papers~\cite{medo2011temporal,wang2013quantifying}, websites~\cite{kong2008experience}, books~\cite{yucesoy2018success}, and patents~\cite{higham2017fame,higham2019ex}, among others.

Surprisingly, we find none of these regularities in the impact of online news. Differently from the widespread heavy-tailed distributions of popularity and impact in social systems, news impact in terms of the number of received comments) is exponentially distributed. Different categories of news have widely different average comment counts, yet their distributions can be collapsed onto a universal exponential distribution. The exponential impact distribution results from the absence or saturation of the widely-studied preferential attachment mechanism. In line with recent findings on the attention decay in science and technology~\cite{higham2017fame,candia2019universal}, the decay of individual news articles follows a universal exponential form. The impact dynamics of online news articles can be reproduced by a parsimonious model with article-level fitness and exponential aging~\cite{golosovsky2018mechanisms}. Building on this model, we can predict the articles' long-term impact based on early activity. We study the impact of natural daily variations of user activity on the dynamics of the article impact and formulate a generalized dynamical model which includes the overall level of user activity as an additional factor along with article fitness and an aging term.

Our findings contribute to the literature on popularity dynamics~\cite{kong2008experience,medo2011temporal,wang2013quantifying, hofman2017prediction,fortunato2018science,yucesoy2018success} by demonstrating that there is a limit to the generality of widely-observed patterns and mechanisms (such as preferential attachment). While previous studies have emphasized the generality of observed patterns of popularity and impact~\cite{candia2019universal}, future research might put more emphasis on identifying violations of pervasive patterns and the causes behind the observed violations. Besides, as managing and influencing the spreading of online information is vital for online newspapers and social platforms, our models and methods can be used to inform decisions by newspaper editors and content creators.

\section*{Results}

\subsection*{News impact is exponentially distributed}
By writing comments, the users demonstrate a higher level of engagement compared to only reading the article~\cite{ksiazek2016user,ksiazek2018commenting}. Importantly, comments are read also by users who do not actively comment, indicating that they play an important role in how a news article is perceived by the public~\cite{barnes2015understanding}. The number of comments can be thus considered as a useful proxy for the article impact~\cite{tsagkias2010news}. To study the distribution of article impact, we discard potential multiple comments from a single user on a given news, thus counting the number of unique users commenting on each article. When all comments are used instead, the results do not change qualitatively (see Supplementary Information, SI, for the results). We further benefit from the additional category information provided directly by the news outlets for all news; the most populated categories are football (BBC) and national (NYT); see Tab.~S1 in SI for details. Since comment counts strongly depend on the category of the news, we analyze news impact individually for each category.

How is the article impact distributed? Impact distributions for creative works are typically found to be heavy-tailed: this is the case for scientific papers~\cite{medo2011temporal}, patents~\cite{valverde2007topology}, and books~\cite{yucesoy2018success}, among others. Broad popularity distributions are also typically found for user-generated content in online systems~\cite{cheng2014can}. Based on these findings, one might expect that the article impact too follows a heavy-tailed distribution. Surprisingly, we find instead that the distributions exhibit exponential tails for both BBC and NYT data. Using the exponential distribution\footnote{Strictly speaking, $P(c)$ should be referred to as a geometric distribution as $c$ is a discrete variable.}
$P(c)\sim\exp(-\lambda\,c)$ for $c\geq c_{\text{min}}$ and following the methodology introduced in~\cite{clauset2009power}, we obtain estimates for the lower bound $\hat{c}_{min}$ and the scaling parameter $\hat{\lambda}$, together with the $p$-value obtained through the Kolmogorov-Smirnov test (see Methods for details).

\begin{figure*}
\centering
\includegraphics[scale=0.75]{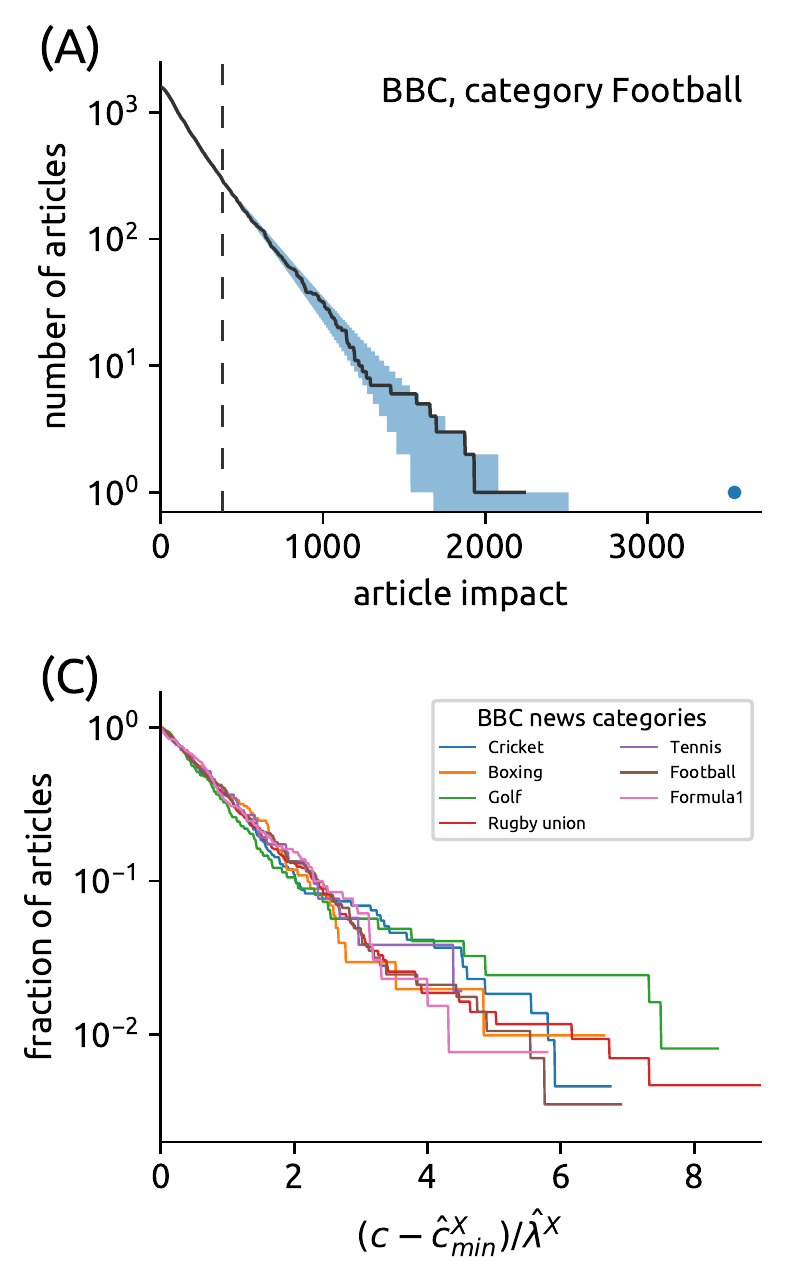}\quad
\includegraphics[scale=0.75]{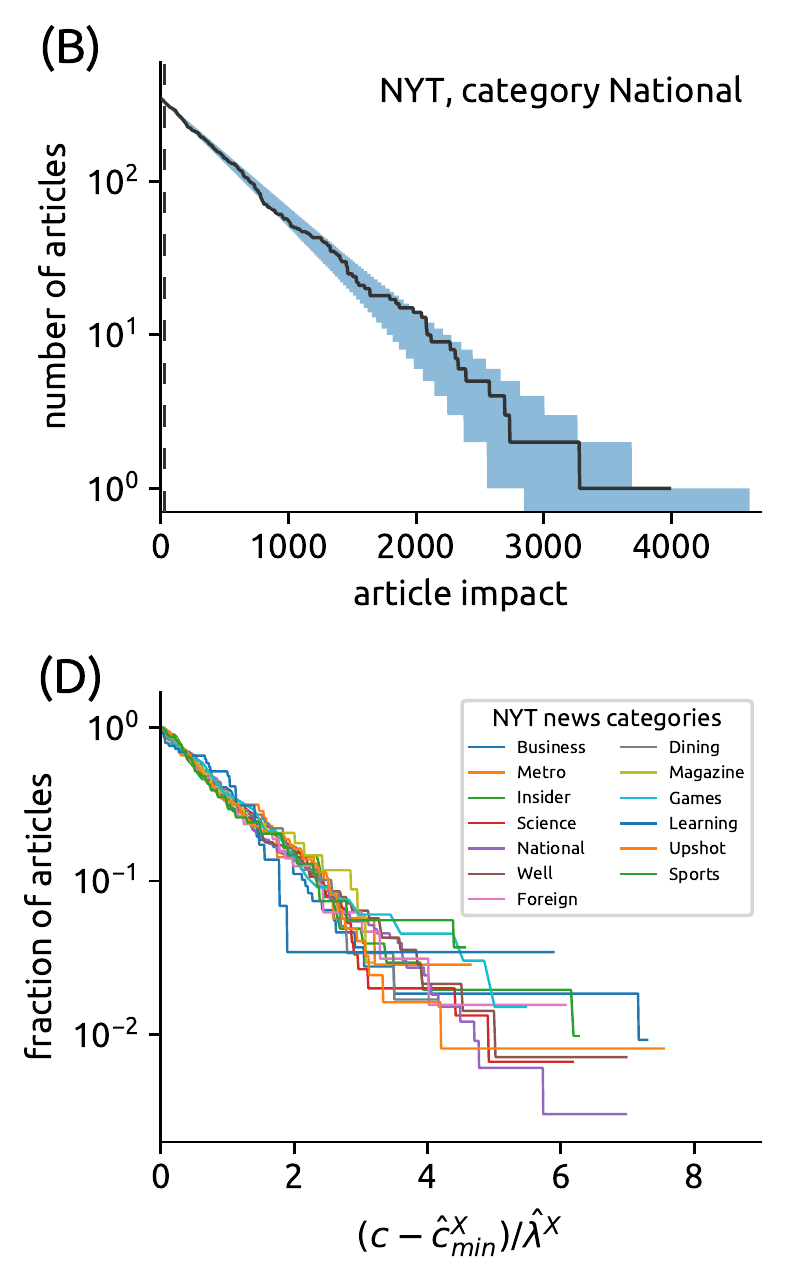}
\caption{\textbf{Article impact is exponentially distributed.} (A, B) Comment count distributions for football news in the BBC data and national news in the NYT data. For the football category, the dot shows a single outlier that was identified in the exponential fitting. (C, D) The distribution of the transformed comment counts, $(c - \cmin^X) / \hat\lambda^X$, in individual news categories; here $\cmin^X$ and $\hat\lambda^X$ are the exponential distribution parameters for category $X$. Upon rescaling, a universal distribution of article impact emerges.}
\label{fig:fig1}
\end{figure*}

For football BBC news, we find that the exponential tail of the distribution begins at $\cmin=381$ and comprises 284 articles (18\% of all football articles). Ignoring a single outler with 3538 comments (see Sec.~S2 in SI for information on outlier detection), the estimated scaling parameter is $\hat\lambda=270\pm17$ and the high $p$-value of $0.96$ indicates that the exponential distribution cannot be ruled out. The good fit can be visually appreciated by observing that the empirical distribution lies within the 5th--95th percentile range of synthetic exponentially distributed data generated with the estimated parameters (Fig.~\ref{fig:fig1}A). For national NYT news, the estimated lower bound is even lower, $\cmin=28$, and the $p$-value is $0.53$ (Fig.~\ref{fig:fig1}C) which again means that an exponential distribution is plausible. Detailed fitting results for all 20 news categories with at least 100 news are shown in SI, Sec.~S2. Importantly, the identified exponential tails are substantial, comprising more than 90\% of news for 10 out of 20 analyzed news categories. The log-likelihood test~\cite{clauset2009power} shows that an exponential distribution fits the data better than a power-law distribution for all categories but one (Learning in the NYT data, see Sec.~S2.1 in SI for additional information).

Inspired by the universality of scientific impact distributions~\cite{radicchi2008universality,fortunato2018science}, we explore an intriguing possibility: By leveraging the estimated parameters, can we collapse the article impact distributions for different categories on top of each other? We find that this is the case: impact distributions in different categories collapse on top of each other after the comment counts are transformed as $(c - \cmin^X) /\hat{\lambda}^X$, where $\cmin^X$ and $\hat{\lambda}^X$ are the estimated lower bound and the scaling parameter for category $X$.

In summary, we find the impact distributions in individual news categories to be far from being power laws. Simple exponential fits work well in several categories where they describe the impact of a majority of news with remarkable veracity.

\subsection*{The relation between article impact and user degree}
The observed article impact distributions, albeit narrower than in many other technosocial systems, still comprise articles that have much greater audience than most other articles. This gives us the possibility to study the relation between impact of an article and degree of the users who have commented on this article. Such a relation can exist if, for example, little active users are mostly idle and comment only on high-impact articles; such a connection would in turn contribute to the high impact of those articles. To assess the level of degree assortativity in the bipartite article-user network, we divide both articles and users in five groups by their degree in such a way that the total degree in each group is approximately the same; groups 1 and 5 have the lowest and the highest degree users/articles, respectively. We then count the numbers of links between respective user and article groups and divide them with the average numbers of links observed in randomized networks. The resulting \emph{link propensity} quantifies how much more likely (if propensity is above one) or less likely (if propensity is below one) are links between a given user and item group compared to a randomized network. For network randomization, we use the recently introduced Dynamic Configuration Model (DCM, \cite{ren2018randomizing}) which is a version of the classical configuration model for networks that grow in time. The DCM internally divides the network in $L$ layers and $L$ is a parameter of the model. Due to the quick aging that we observe in the analyzed datasets (see the following sections), we chose the number of layers to be the same as the number of days in each respective dataset (the results are robust with respect to the choice of $L$). See \cite{medo2019optimal} for a principled way to determine $L$ in a monopartite growing network.

\begin{figure*}
\centering
\includegraphics[scale=0.9]{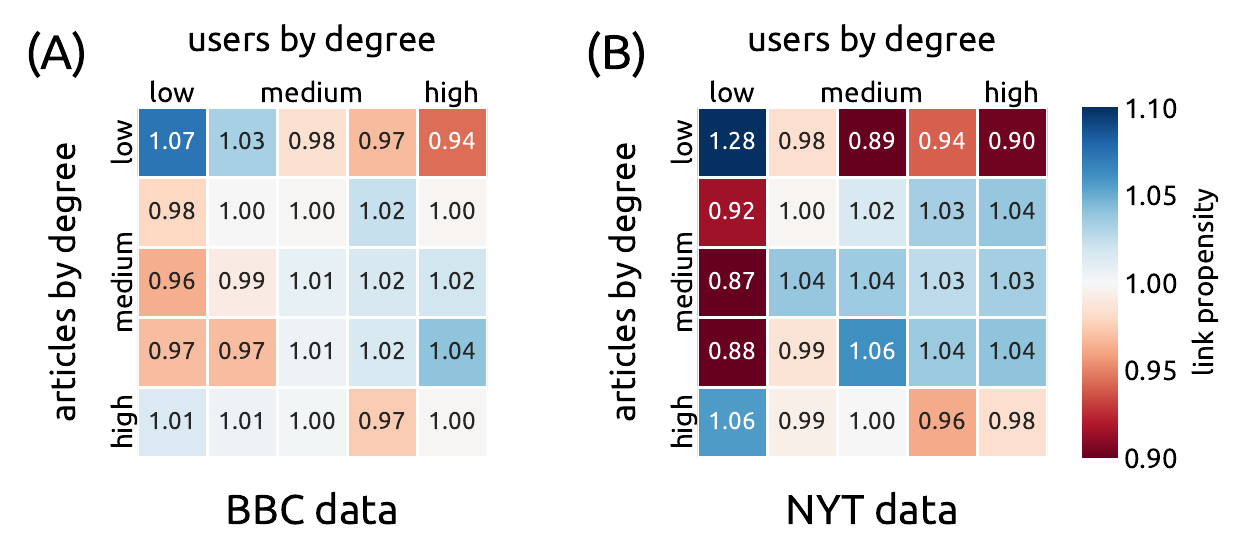}
\caption{\textbf{Relative link propensity between users and articles of various degree.} Both users and articles are divided in five groups by their degree. The relative link propensity values quantify the excess (values greater than one) or lack (values smaller than one) of links between the respective pair of user- and article-groups as compared to randomized networks (we average over 1,000 realizations of the DCM model~\cite{ren2018randomizing}). The left and right panel show the results for the BBC data and the NYT data, respectively. While some statistically significant deviations from the null model can be observed (values that differ from one by more than, approximately, 0.02 have absolute $z$-scores above 3), only one of them (links between low-degree users and low-degree articles in the NYT data) is larger than 20\% either way.}
\label{fig:SI_link_propensity}
\end{figure*}

Figure~\ref{fig:SI_link_propensity} shows that the relative link propensity is close to one for most pairs of user- and article-groups. One emerging pattern shared by the BBC and NYT data concerns the least popular articles which are commented by the least active users more than expected and by the most active users less than expected. By contrast, the original hypothesis of the most popular articles owing their popularity to little active users is ruled out by the results. For the BBC data, the relative link propensity between the most popular articles and the least active users does not differ significantly from one. For the NYT data, it is only 1.06 which means that the most popular articles do not receive 20\% of their comments from the group of least active users who write 20\% of all comments but $20\%\times1.06\approx 21\%$ which is a negligible increase. We can conclude that the most popular articles receive comments from users of all degree values approximately in line with expectations.

\subsection*{Preferential attachment plays a minor role in the dynamics of impact}
The empirical exponential distributions of article impact inevitably lead us to investigate possible mechanisms behind their emergence. Motivated by existing results on the dynamics of impact for cultural products as diverse as scientific papers~\cite{medo2011temporal,wang2013quantifying}, patents~\cite{higham2017fame}, and bestseller books~\cite{yucesoy2018success}, one expects two main forces shaping the dynamics of news impact~\cite{candia2019universal}: preferential attachment and temporal decay. We start by addressing preferential attachment which implies that the rate at which article $i$ receives new comments, $\Delta c_i(t)/\Delta t$ where $\Delta c_i(t) = c_i(t + \Delta t) - c_i(t)$, is a power-law function (most commonly, a linear function) of the number of already-received comments, $c_i(t)$.

\begin{figure*}
\centering
\includegraphics[scale=0.75]{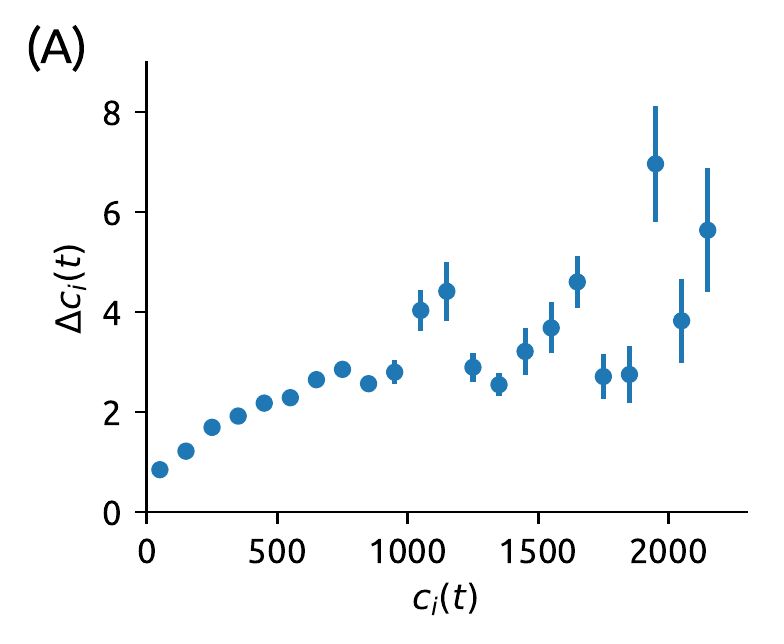}
\includegraphics[scale=0.75]{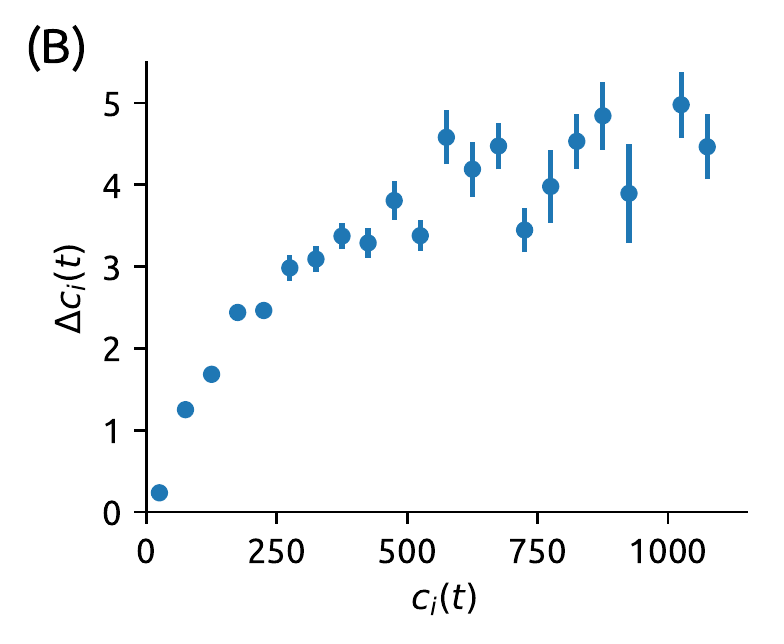}
\caption{\textbf{Preferential attachment in the dynamics of article impact.}  The number of new comments in $\Delta t=10\,\text{min}$ as a function of the current number of comments. For the BBC data (A), the fit up to the comment count 800 yields the slowly-growing dependence proportional to $1+c_i(t)/220$. Above 800 comments, the dependence is even weaker (saturation). For the NYT data (B), sublinear preferential attachment with the exponent $0.79$ is the best fit, followed by saturation for $c_i(t)\gtrsim400$. Error bars indicate the standard error of the mean.}
\label{fig:fig2}
\end{figure*}

In contrast with pervasive findings in the popularity dynamics literature, we find that preferential attachment is negligible in the BBC data (Fig.~\ref{fig:fig2}A) and exhibits clear sub-linearity and saturation for the NYT data (Fig.~\ref{fig:fig2}B). More specifically, in the BBC data, more than 200 comments are needed for an article to double its commenting rate with respect to a comment-free article. Furthermore, the observed weak growth of $\Delta c_i(t)$ with $c_i(t)$ can be explained in terms of a dynamic model where no preferential attachment is present (see Fig.~S9 in SI). In the NYT data, $\Delta c_i(t)$ first grows as a power of $c_i(t)$ with an exponent below one (sub-linear preferential attachment~\cite{krapivsky2000connectivity}, see Fig.~S10 in SI) and becomes independent of $c_i(t)$ for $c_i(t)\gtrsim 400$. The lack of/saturation of preferential attachment has important consequences as it prevents a power-law degree distribution from emerging (see Section on modeling the article impact dynamics and Sec.~S4 in SI).
In summary, we find that despite the articles' comment counts explicitly reported by both BBC and NYT (see Fig.~S1 in SI), the impact of preferential attachment on the dynamics of news article impact is limited.

\subsection*{The dynamics of article impact follows an exponential decay}
\begin{figure*}
\centering
\includegraphics[scale=0.75]{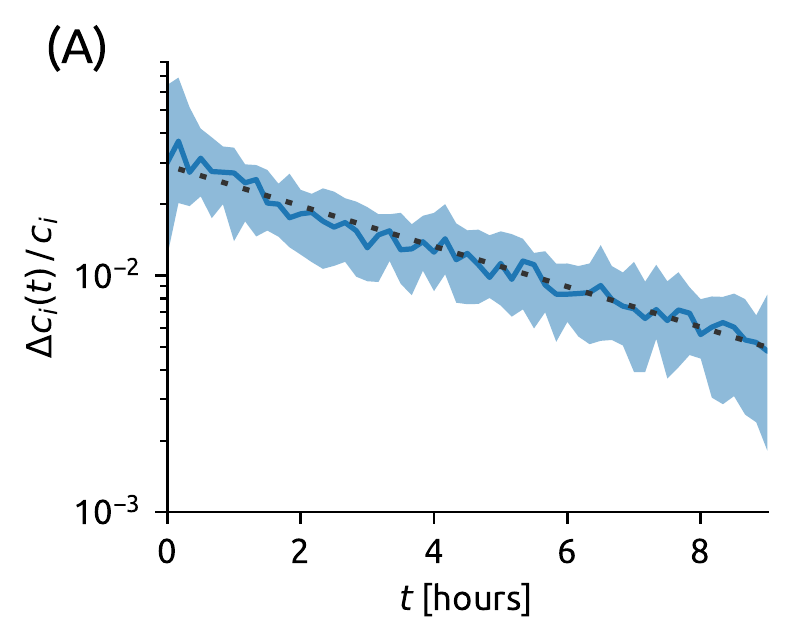}
\includegraphics[scale=0.75]{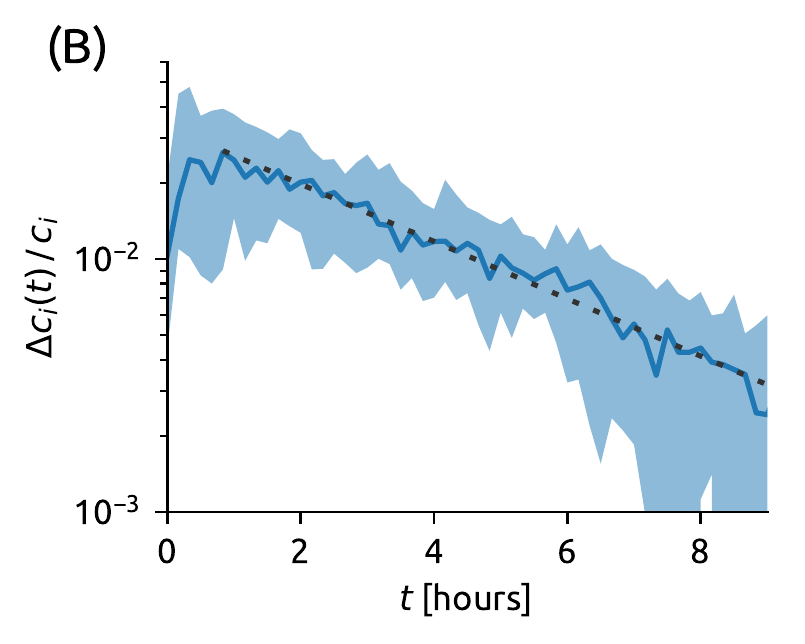}
\caption{\textbf{Aging in the dynamics of article impact.} The number of new comments of an article, $\Delta c_i(t)$, normalized by the final number of comments, $c_i$, as a function of its age, $t$, for the hit articles (90th percentile by the number of comments). The dotted lines indicate the linear fit for age 0--10 hours; their slopes correspond to representative aging timescales $\varTheta=305\,\text{min}$ (BBC, panel A) and $\varTheta=230\,\text{min}$ (NYT, panel B), respectively.
The time-of-day effects are suppressed here by including only the articles that appear in the morning between 9am and noon (BBC) and between 2pm and 5pm GMT (NYT). The shaded areas there indicate the 20th--80th percentile range and the solid lines show the median values for the considered articles.}
\label{fig:fig3}
\end{figure*}

Existing studies have found various functional forms for the decay of the impact of cultural items, including power-law~\cite{crane2008robust}, log-normal~\cite{wang2013quantifying,yucesoy2018success}, exponential~\cite{parolo2015attention,higham2017fame}, stretched exponential~\cite{wu2007novelty}, and biexponential~\cite{candia2019universal}. To quantify the temporal decay of article impact, for each news $i$, we measure the news' number of new comments relative to the article's final comment count, $f_i(t):=\Delta c_i(t)/c_i$, as a function of the article age, $t$. The normalization by the article's final comment count makes the dynamics of articles of different ultimate impact directly comparable.\footnote{This normalization is also motivated by the observation that under negligible preferential attachment, $\Delta c_i(t)/c_i$ is expected to accurately capture the aging function of the articles, as shown in the next section.}
For each age, $t$, we compute the median of $\Delta c_i(t)/c_i$ over all considered articles, obtaining the representative decay function, $f(t)$. We restrict the analysis to \emph{hit articles} which, for the purpose of this work, are defined as the articles whose number of comments is above the 90th percentile (679 and 428 comments in BBC and NYT, respectively). To suppress the time-of-day effects, we include only the BBC articles that appeared in the morning between 9am and noon---the 10 hour range shown in Fig.~\ref{fig:fig3}A is thus a period when user activity is rather uniform at the BBC website. User activity is substantially lower in the night, which directly effects the evolution of $c_i(t)$ (see Sec.~S3 in SI for more details). For the same reasons, we focus on the NYT articles that appeared between 2pm and 5pm GMT.

We find that the articles' temporal decay follows a universal exponential form (Figs.~\ref{fig:fig3}). In particular, the average decay function $f(t)$ can be accurately fitted by an exponential function: $f(t)=\ee^{-t/\varTheta}$ where $\varTheta=305\,\text{min}$ for BBC and $\varTheta=230\,\text{min}$ for NYT. While $f(t)$ decreases exponentially in the BBC data during the whole observed range, it shows a short period (approximately 1 hour) of increase in the NYT data. This is a direct consequence of the preferential attachment that applies for low comment counts---as the number of comments grows, the rate of commenting initially accelerates before aging in combination with sublinear/saturated preferential attachment eventually cause the rate of commenting to decrease.

Our finding of a regular exponential decay of article impact agrees with the report of real news on Twitter differing from rumors by exhibiting a regular pattern of a monotonous decrease of attention~\cite{kwon2013prominent}. The observed exponential aging can be explained by a simple model where each article is of interest to a fixed pool of readers (the pool's size is determined by the article's attractiveness to the readers) and every reader has a fixed probability to write their comment (most readers comment in a discussion only once) per time unit~\cite{ishii2012hit}. The observed exponential decay can be also interpreted as a limit scenario of the bi-exponential impact decay predicted by a recent work based on a model with communication memory and cultural memory~\cite{candia2019universal}. The reason why such a limit scenario holds for online news needs to be clarified by future research. A plausible hypothesis is that as the comments to online news articles unfold over a narrow time period following a news, we cannot use them to observe the process whereby the communication memory associated with an article is converted into cultural memory. If this is the case, the model in~\cite{candia2019universal} predicts an exponential decay of collective attention, in line with our observed decay functions.

\subsection*{Exponentially-distributed fitness and exponential aging shape the dynamics of article impact}
The impact dynamics for scientific papers~\cite{medo2011temporal,wang2013quantifying} and bestseller books~\cite{yucesoy2018success} is typically modeled in terms of preferential attachment, fitness and aging. Building on these studies, a potential model for the commenting dynamics would assume that the expected rate at which article $i$ receives new comments at time $t$ is
\begin{equation}
\label{PFA_model}
\Delta c_i(t)/\Delta t = [1+c_i(t)]\,\eta_i\,f_i(t-t_i)
\end{equation}
where $1+c_i(t)$ is the preferential attachment factor, $\eta_i$ is the fitness factor, $f_i(t-t_i)$ denotes an article-dependent aging function, and $t_i$ is the appearance time of article $i$. In line with previous studies~\cite{medo2011temporal,wang2013quantifying}, article fitness $\eta$ is a hidden intrinsic parameter that quantifies, other factors being equal, how a given article is attractive to the website's audience. We refer to this model as the PFA model because it includes Preferential attachment, Fitness and Aging. In this model, a narrow exponential distribution of article fitness, $\rho(\eta)=\exp{(-\eta)}$, leads to the emergence of a power-law distribution of the comment count~\cite{medo2011temporal}. In other words, small differences in items' fitness are amplified by preferential attachment and produce wide impact inequalities.

\begin{figure*}
\centering
\includegraphics[scale=0.75]{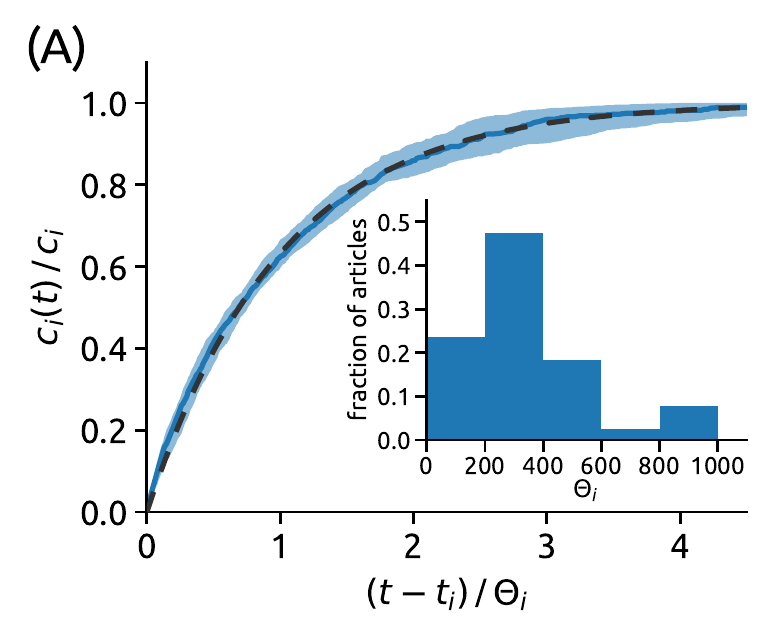}
\includegraphics[scale=0.75]{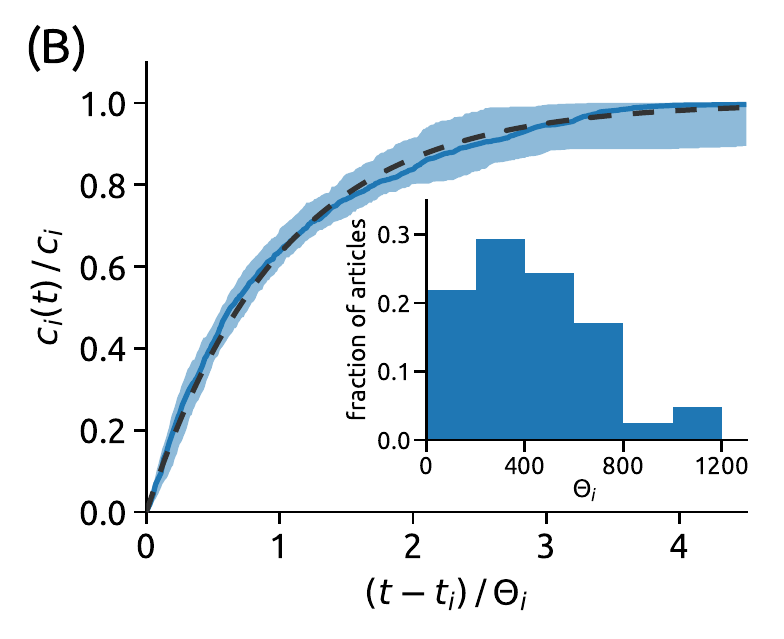}
\caption{\textbf{The universal dynamics of article impact.} The comment count evolution in terms of the normalized article age $(t-t_i)/\Theta_i$ for the BBC (A) and NYT (B) data. The shaded areas there indicate the 20th--80th percentile range and the solid lines show the median values for the considered articles. The dashed line represents the proposed model and its solution given by \eref{law}. The inset shows the distribution of the timescales $\Theta_i$ obtained by minimizing the Kolmogorov-Smirnov statistic. Only ``morning'' articles are included in the analysis, as in Figue~\ref{fig:fig3}.}
\label{fig:fig5}
\end{figure*}

The observed weak preferential attachment and exponential temporal decay suggest a simpler model of the dynamics of article impact where only article fitness and exponential aging play a role. We thus assume that the rate at which article $i$ receives new comments at time $t$ is
\begin{equation}
\label{FA_model}
\Delta c_i(t)/\Delta t = \eta_i\,f_i(t-t_i)
\end{equation}
which we refer as the FA (Fitness-Aging) model~\cite{golosovsky2018mechanisms}. To accurately represent the commenting dynamics, we introduce individual aging timescales $\Theta_i$ and the aging factor in the form $f_i(t-t_i)=\exp[-(t-t_i)/\Theta_i]$. The aging timescales $\Theta_i$ are estimated from the empirical data by minimizing the Kolmogorov-Smirnov statistic between the comment count dynamics in the model and in the empirical data (see Sec.~S6 in SI). If $\Theta_i\gg1$, the expected final comment count under the FA model is directly proportional to the product of the article fitness and the aging timescale, $\overline{c_i}=\eta_i\Theta_i$ (see Sec.~S5 in SI). The model further implies that
\begin{equation}
\label{law}
\frac{\overline{c_i(t)}}{\overline{c_i}} = 1 - \exp\biggl(-\frac{t-t_i}{\Theta_i}\biggr).
\end{equation}
Motivated by this result, we measure the dynamics of the comment count normalized by the final comment count. We find that \eref{law} captures the empirical dynamics remarkably well (Fig.~\ref{fig:fig5}) and allows us to collapse all article trajectories onto a universal curve (Figure~\ref{fig:fig5}). This result demonstrates that the fitness-aging model captures the two essential factors that govern the dynamics of news article impact, and it further confirms that preferential attachment plays a negligible role in the emergence of hit articles.

Since $\overline{c_i}\sim\eta_i\Theta_i$, exponentially distributed $\eta\Theta$ leads to the emergence of an exponential comment count distribution in line with the empirical data. When the aging timescales vary relatively little among the articles, as is the case here, the distribution of article fitness alone is approximately exponential. Interestingly, an exponential distribution of $\eta\Theta$ (referred to as total relevance therein) was reported in~\cite{medo2011temporal} for scientific papers and an exponential distribution of $\eta$ (in a model without aging) was reported in~\cite{kong2008experience} for pages of the World Wide Web. To identify theoretical mechanisms behind this widespread emergence of exponentially distributed fitness of items remains an important future challenge.

\subsection*{Early activity can be used to predict article impact}
The regular dynamics demonstrated by Fig.~\ref{fig:fig5} suggests that the early commenting activity and the final article impact are highly correlated. To verify this conjecture, we study a classification problem where we aim to predict whether an article will become a hit (\ie, if it will belong to the 90th percentile by the final impact). We classify an article as positive if it belongs to the 90th percentile by the number of comments that it has attracted over the first $\Delta t$ minutes, and negative otherwise. We evaluate the classifier using precision and AUC which are both classical information retrieval metrics~\cite{manning2010introduction} that range from zero (the worst result) to one (the best result). We find that the proposed simple classifier exhibits high values of precision and AUC even when $\Delta t$ is short: precision exceeds 0.6 after five minutes, for example (see Table~\ref{tab:classification} for full results).

The observed predictability is unsurprising given previous results on the correlation between early and late popularity of online content~\cite{szabo2010predicting,cheng2014can,shulman2016predictability}. However, previous studies interpreted the early-stage predictability of the virality of online cascades as a possible manifestation of cumulative advantage~\cite{shulman2016predictability}. This cannot be the case here for online news where we demonstrated that preferential attachment has a negligible effect. Taken together, our findings suggest a somewhat simpler scenario: the news that are highly attractive for the public tend to receive more connections throughout their whole lifetime than less attractive news. In this sense, the impact of online news might be seen as more ``meritocratic'' than that of content in systems with preferential attachment: The news with truly high fitness are those that eventually succeed, regardless of cumulative advantage effects.

\begin{table}
\centering
\setlength{\tabcolsep}{12pt}
\begin{tabular}{rrrrr}
\hline
& \multicolumn{2}{c}{BBC data} & \multicolumn{2}{c}{NYT data}\\
$\Delta t$ & $P$ & $AUC$ & $P$ & $AUC$\\
\hline
   1 & 0.33 & 0.69 & 0.31 & 0.62\\
   2 & 0.51 & 0.81 & 0.51 & 0.75\\
   5 & 0.60 & 0.87 & 0.64 & 0.89\\
  10 & 0.63 & 0.89 & 0.73 & 0.92\\
  60 & 0.69 & 0.92 & 0.81 & 0.96\\
 240 & 0.77 & 0.94 & 0.83 & 0.98\\
1200 & 0.93 & 0.99 & 0.95 & 0.99\\
\hline
\end{tabular}
\vspace*{8pt}
\caption{Classification precision and AUC for the hit articles.}
\label{tab:classification}
\end{table}

\subsection*{Circadian patterns of user activity patterns shape the dynamics of news impact}
\begin{figure*}
\centering
\includegraphics[scale=0.69]{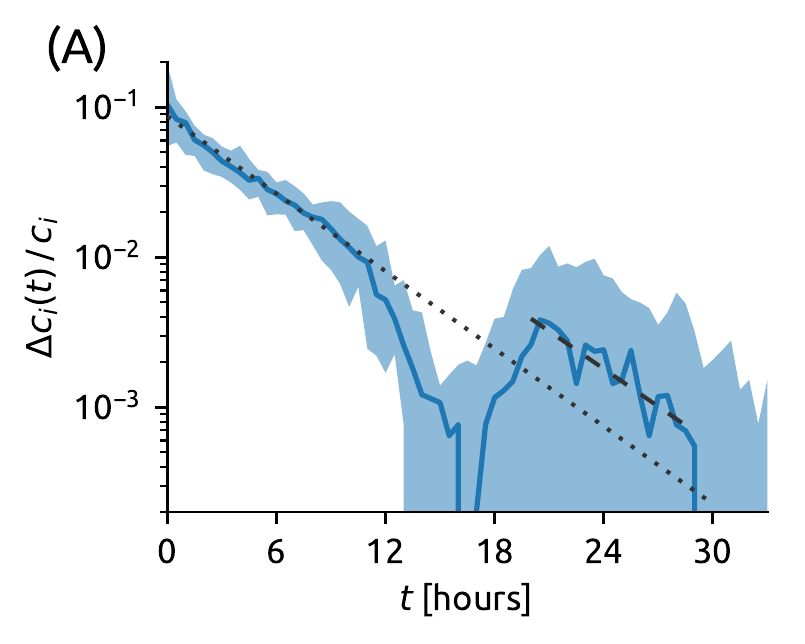}
\includegraphics[scale=0.69]{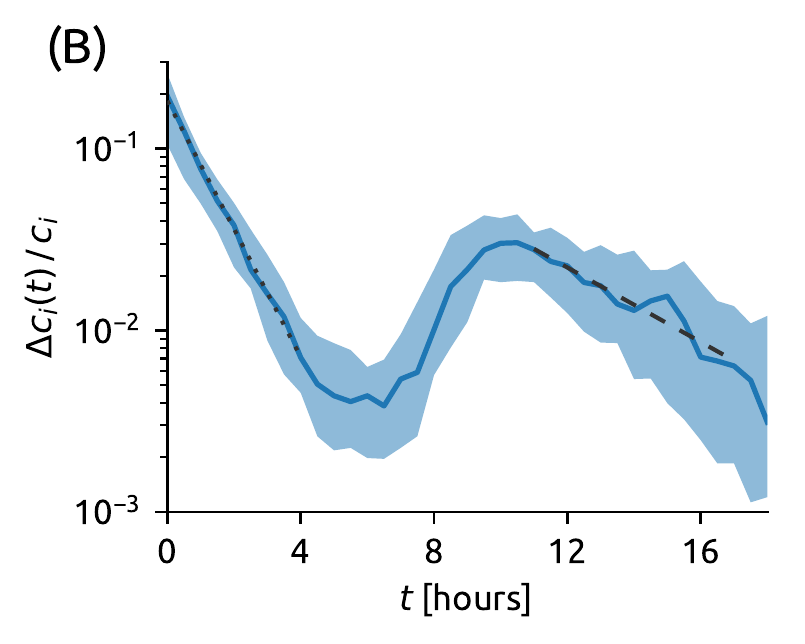}
\includegraphics[scale=0.69]{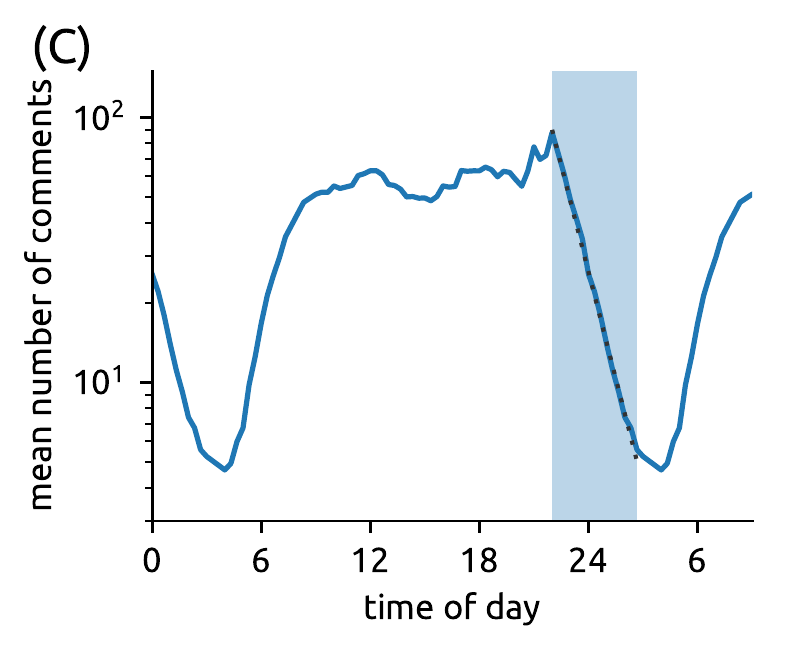}
\caption{\textbf{Interplay between exponential aging and circadian user activity patterns in the BBC data.} (A) The number of new comments of an article, $\Delta c_i(t,\Delta t)$, normalized by the final number of comments, $c_i$, as a function of its age, $t$, for morning hit articles (published between 9 am and noon). The dotted line indicates the linear fit for age 0--10 hours (fitted timescale 302 minutes). The dashed line indicates the linear fit for age 20--29 hours (fitted timescale 312 minutes). (B) As (A) for evening articles (published between 9 pm and midnight). Timescales of the indicated fits are 74 minutes (first 4 hours, dotted line) and 254 minutes (age 11--17 hours, dashed line). In panels (A) and (B), the age bin size is 30 minutes to achieve better statistics for high article age. (C) The course of the mean number of comments in 20 minute intervals during the day at the BBC Sport website. The fitted timescale of the exponential decrease between 10 pm and 3 am is 97 minutes.}
\label{fig:fig6}
\end{figure*}

To study the impact dynamics, we have until now focused specifically on ``morning'' articles that benefit from high and approximately constant user activity for more than ten hours after their publication. We now turn our attention to the effect of overall user activity, which naturally decreases in the night (see Fig.~S7 in SI), on the dynamics of article impact. Figure~\ref{fig:fig6}A shows the evolution of the median number of new comments for the same set of morning articles over a longer period. This allows us to observe a decrease of commenting activity in the night (age 12--18 hours) and renewed exponential decay on the second day (age 20--29 hours) with the timescale of 312 minutes. The two fitted exponential decay timescales, 302 minutes for article age 0--10 hours\footnote{The minor difference with respect to Figure~\ref{fig:fig3} where the fitted timescale is 305 minutes is due to the use of wider bins in Figure~\ref{fig:fig6}A in order to improve the statistics for high article age.}
and 312 minutes for article age 20--29 hours are remarkably close to each other. We see that after user activity recommences after a night, article aging continues in the same speed than before the night.

Figure~\ref{fig:fig6}B shows the evolution of the median number of new comments for evening hit articles (articles published between 9 pm and midnight). We see again two phases of exponential decay: an early phase during the night (age up to 4 hours) with the timescale of 74 minutes and a late phase during the day (age 11--17 hours) with the timescale 254 minutes. Albeit having a somewhat shorter timescale, the late phase is a direct equivalent of the previously observed aging of morning articles. We thus see that morning and evening articles exhibit similar aging during the day.

On the other hand, the early night aging of evening articles has a timescale which is significantly shorter than the aging timescle observed during the day. To better understand this fast aging, Figure~\ref{fig:fig6}C shows the average overall activity on the BBC Sport website. We see that after an approximately constant activity during the day (from 8 am until 8 pm) and a small peak in the evening (from 8 pm until 10 pm) when many sport events take place, user activity dramatically decreases after 10 pm. Moreover, the user activity decay during the first five hours of the decrease (from 10 pm until 3 am) is a nearly perfect exponential with the fitted timescale of 97 minutes. Varying user activity can be included in the previously introduced FA model described by \eref{FA_model} by introducing it as an additional multiplicative factor, thus obtaining an FAA (Fitness-Aging-Activity) model. The rate at which article $i$ receives new comments at time $t$ then reads
\begin{equation}
\label{FAA_model}
\Delta c_i(t) / \Delta t = \eta_i\,f_i(t-t_i)\,A(t)
\end{equation}
where $t_i$ and $\eta_i$ are the appearance time and fitness of article $i$, respectively, and $A(t)$ is the overall activity factor which is common to all articles at the platform. If we now assume that article $i$ appears when user activity, $A(t)$, decreases exponentially, we obtain
$$
\Delta c_i(t) / \Delta t = \eta_i\exp[-(t-t_i) / \Theta_A] A(t_i) \exp[-(t - t_i)/\Theta_U]
$$
where $\Theta_A$ and $\Theta_U$ are the article and user exponential decay timescales, respectively. The two exponential terms can be combined in one as
$$
\exp[-(t-t_i) / \Theta_A] \times \exp[-(t - t_i)/\Theta_U] = \exp[-(t - t_i)/\Theta_J]
$$
where the joint exponential timescale, $\Theta_J$, has the form
\begin{equation}
\label{joint_timescale}
1/\Theta_J = 1/\Theta_A+1/\Theta_U.
\end{equation}
Using the fitted values $\Theta_A=302\,\text{min}$ and $\Theta_U=97\,\text{min}$, we obtain $\Theta_J=73\,\text{min}$ which is in an excellent agreement with the fitted aging timescale 74 minutes of evening articles during the night. Results are qualitatively similar for the NYT data (see Sec.~S7 in SI). We can thus conclude that the FAA model given by \eref{FAA_model} presents an effective way of combining article dynamics with circadian and other patterns of varying user activity.

\section*{Discussion}
By analyzing data on the comments to online news articles in two major nationwide newspapers, we were able to uncover surprising empirical regularities that characterize the distribution of the impact of online news articles and the impact dynamics. In particular, we revealed two universal patterns: (1) For both newspapers, the distribution of the number of comments received by articles from various categories collapse onto a universal exponential curve and (2) the dynamics of the comment count of different news articles collapse onto a universal curve once appropriate rescaling is applied. The exponential impact distribution emerges from impact dynamics where preferential attachment plays a negligible role. This indicates that differently from other social systems~\cite{salganik2006experimental,wu2007novelty}, popularity signals in online newspapers are not prominent enough to play a significant role. Main empirical dynamical patterns of article impact can be reproduced with a minimal model which combines article fitness, an aging term (which in our case has an exponential form), and overall user activity. When user activity is approximately constant, only article impact and aging remain and the resulting dynamics is particularly simple.

Our findings contrast with the previous literature on success and popularity that has emphasized that success and popularity are usually characterized by heavy-tailed distributions~\cite{kong2008experience,radicchi2008universality,cheng2014can,broido2019scale}, and that preferential attachment plays a key role in shaping the emergence of hits~\cite{kong2008experience,medo2011temporal,wang2013quantifying,yucesoy2018success}. Additional research is needed to quantify the relative importance of different factors that trigger user engagement in a news article (\ie, which article attributes contribute to its fitness~\cite{berger2012makes,bandari2012pulse}), and how our findings generalize to different cultures and platforms in languages other than English.

We quantified the impact of a news article through the number of comments it received from the online newspaper's readers. Other metrics of impact might be also relevant to news outlets. For example, the overall impact of a news can be quantified as a combination of the impact on the readers of the newspaper and the impact on users who shared or commented the news in different social media and news aggregation platforms. Uncovering the regularities of the news articles' dynamics by incorporating data from social media and news aggregators is an important direction for future research, given the critical role of these platforms for news dissemination~\cite{hermida2012share,dellarocas2015attention}.

Although our study focused on news outlets that only include verified news (BBC and NYT), our findings can inspire future studies related to the spreading of misinformation in online systems~\cite{del2016spreading,vosoughi2018spread}. Our results could serve as baselines in future studies that consider the commenting dynamics of both verified and false news. Do false news trigger different patterns of impact compared to true news? Is the diffusion of false and true news governed by different fundamental mechanisms? Understanding which mechanisms play a major role in engaging users and triggering their comments might suggest intervention strategies to prevent their impact.

The collected BBC data contain several other characteristics that have not been included in the present study: comment length, comment text, as well as the number of up-votes and down-votes for each comment. Their analysis can yield further patterns in article commenting. Of particular interest is the interplay between comment sentiment and the discussion activity is of particular interest. Do positive or controversial comments help fuel the discussion? Do early emotionally loaded comments influence the long-term tone of the discussion? Which factors contribute to users approval or disapproval of a comment? Such studies can help us understand how we discuss online and how to make these discussions more construcitve.

\section*{Methods}

\subsection*{Empirical datatasets}
We regularly crawled the sport section of the BBC website (its front page and the pages dedicated to individual sports) and collected the found news articles with commenting sections. From October 1, 2018 to June 30, 2019, we collected 3,087 articles that received 852,400 comments from 67,527 readers. Each article is assigned to a sport category. The most populated categories are Football (1590 articles), Rugby Union (439 articles), Cricket (240 articles), Tennis (162 articles), Formula 1 (139 articles), Golf (123 articles) and Boxing (103 articles). Each comment is time-stamped with the time resolution of one minute. BBC typically closes commenting on the second midnight after the article has been published; most of them are therefore open for 24--48 hours.

We complement the unique BBC dataset with a dataset containing articles with commenting sections from the New York Times (NYT).\footnote{Data obtained from \protect\url{https://www.kaggle.com/aashita/nyt-comments}.}
From January 1, 2017 to May 30, 2017, there are 2,801 articles that received 649,794 comments from 75,118 readers. Also here, each article is assigned to a category. Unlike for BBC, sport articles are a minority in the NYT data: The most populated categories are National (348 articles), Learning (306 articles), Magazine (262 articles), Sports (213 articles) and Foreign (204 articles). Each comment is time-stamped with the time resolution of one minute. While some comments arrive long after the articles are published, the median time after which the hit articles (90th percentile by the comment count) receive 99\% of their comments is less than 26 hours. To study the article impact dynamics, we thus focus on the first 26 hours of article age (the final article impact is nevertheless determined using all data). See Supplementary Information, Section~S1, for detailed information about the datasets.

\subsection*{Fitting the comment count distributions}
The maximum likelihood estimate (MLE) of the scaling parameter of the exponential distribution is known to be the sample mean, $\hat\lambda = (\sum_{i=1}^n c_i) / n$. As can be seen from Figure~\ref{fig:fig1}, the comment count distribution follows an exponential form starting from some lower bound $\cmin$. The MLE estimate then changes to
$\hat\lambda(\cmin) = [\sum_j (c_j - \cmin)] / n(\cmin)$ where the summation is over $j$ for which $c_j \geq\cmin$ and $n(\cmin) = \abs{\{j:\ c_j \geq\cmin\}}$ is the number of comment counts that match or exceed the lower bound. We assess the estimate uncertainty using non-parametric bootstrap---standard deviation of the MLE estimates is evaluated for 10,000 bootstrap realizations of the comment count data.

To determine $\cmin$, we follow the approach suggested by \cite{clauset2009power}: We choose $\cmin$ that minimizes the difference between the comment count distribution and the fitted exponential distribution as measured by the standard Kolmogorov-Smirnov statistic which has the form
\begin{equation}
\label{KS}
D = \max_{c\geq\cmin} \abs{S(c) - P(c)}
\end{equation}
where $S(c)$ and $P(c)$ are the cumulative distributions for the comment counts and the fitted exponential distribution, respectively. When the weighted Kolmogorov-Smirnov statistic~\cite{clauset2009power} is used, which puts more weight on tails of the distributions, results do not change qualitatively. This further suggests that our fitting procedure and the conclusions drawn from the results are robust.

The next step is to test the hypothesis that the observed comment counts indeed follow an exponential distribution. We follow again \cite{clauset2009power} where the authors suggest to use the fitted parameters to generate synthetic exponentially distributed datasets, fit each of those datasets as described above, and finally calculate the $p$-value as the fraction of synthetic datasets whose resulting $D$ exceeds that obtained for the real data.

To finally compare the statistical evidence for an exponential distribution with that for a power-law distribution, we do the same analysis for fitting a power-law distribution. Since the input data are discrete, the MLE cannot be given in a closed form \cite{clauset2009power}, we numerically maximize the log-likelihood
\begin{equation}
\mathcal{L}(\alpha,\cmin) = -n(\cmin)\zeta(\alpha,\cmin) -
\alpha\sum_{j:\ c_j\geq\cmin}\ln c_j.
\end{equation}
A detailed comparison between fitting exponential and power-law distribution to the commenting data, including the log-likelihood test which directly compares the likelihood that the analyzed data has been drawn from the exponential or the power-law distribution, is presented in Sec.~S2 in SI.

\subsection*{Abbreviations}
\noindent BBC: British Broadcasting Corporation

\noindent NYT: New York Times

\noindent SI: Supplementary Information

\noindent DCM: Dynamic Configuration Model

\noindent PFA: Preferential Attachment-Fitness-Aging (model)

\noindent FA: Fitness-Aging (model)

\noindent AUC: Area Under the receiver operating Characteristic

\noindent FAA: Fitness-Aging-Activity (model)

\noindent MLE: Maximum Likelihood Estimate

\subsection*{Availability of data and materials}
Upon publication of the manuscript, the BBC commenting data and scripts to reproduce the results presented here will be made available at \url{https://github.com/8medom/Article-Impact}.

\begin{acknowledgments}
This work is supported by the National Natural Science Foundation of China (Grant Nos.~11622538, 61673150, 11850410444). MSM acknowledges financial support from the URPP Social Networks at the University of Zurich, the Swiss National Science Foundation (Grant No. 200021-182659), and the UESTC professor research start-up (Grant No.~ZYGX2018KYQD215). LL acknowledges the Science Strength Promotion Programme of UESTC. We thank Patrick Park for stimulating discussions.
\end{acknowledgments}

\bibliography{news_references}

\clearpage  

\onecolumngrid
\renewcommand\thefigure{S\arabic{figure}}
\setcounter{figure}{0}
\renewcommand{\thesection}{S\arabic{section}}
\setcounter{section}{0}
\renewcommand{\thesubsection}{S\arabic{section}.\arabic{subsection}}
\setcounter{subsection}{0}
\renewcommand{\theequation}{S\arabic{equation}}
\setcounter{equation}{0}

\begin{center}
\LARGE Supporting Information
\end{center}

\section{Data description}
\label{sec:data_description}

\subsection{The BBC article discussion data}
We collected a comprehensive dataset of sport news articles with discussions by periodically crawling the BBC Sport website (its front page and the pages dedicated to individual sports). In the time period from October 1, 2018 until June 30, 2019 (273 days), there were 3,087 article discussions open that received 852,400 comments from 67,527 users. The user median and mean number of comments are 2 and 12.6, respectively. The median and mean number of comments of a news article are 155 and 276, respectively. We measure the news article impact by the number of unique users who left a comment in its discussion. The median and mean article impact are 108 and 180, respectively.

Each comment is time-stamped with the time resolution of one minute. BBC typically closes discussions on the second midnight after the corresponding news article has been published; most of them are therefore open for 24--48 hours. There are two exceptions in the dataset: one discussion that has been open marginally longer than 48 hours and another discussion that was open for 13 days; it attracted only a few comments after day two, though.

\subsection{The NYT article discussion data}
We further support our findings using the New York Times (NYT) commenting data obtained from \url{https://www.kaggle.com/aashita/nyt-comments}. Our NYT dataset comprises articles published in January--May 2017. At the NYT, it is possible to comment on a previously written comment (in fact, several levels of response are possible). To measure the article impact, we consider only the top-level comments; responses to comments are neglected as they are driven by the comments to which the responses are made. There are 2,801 articles and 649,794 comments from 75,118 users. The user median and mean number of comments are 1 and 6.0, respectively. The median and mean number of comments of a news article are 38 and 165, respectively. We measure the news article impact by the number of unique users who left a comment in its a discussion. The median and mean article impact are 37 and 143, respectively.

Each comment is time-stamped with the time resolution of one minute. Unlike the BBC data, article discussions at the NYT remain open for long time. Despite this, the article commenting activity decays quickly and the median time to reach 99\% of the final comment count is approximately 26 hours (taking into account the hit articles whose comment count is above the 90th percentile).

\begin{table*}
\centering
\begin{tabular}{lrrr}
\hline
Category    & Articles & Mean comment count & Mean number of unique users (impact)\\
\hline
Football    & 1590 & 361 & 236\\
Rugby Union & 439  & 219 & 130\\
Cricket     & 240  & 213 & 133\\
Tennis      & 162  & 157 & 108\\
Formula 1   & 139  & 338 & 209\\
Golf        & 123  & 117 &  85\\
Boxing      & 103  & 195 & 152\\
\hline
National    & 348 & 678 & 563\\
Learning    & 306 &  59 & 52\\
Magazine    & 262 &  77 & 72\\
Sports      & 213 &  48 & 45\\
Foreign     & 204 & 277 & 241\\
Games       & 192 &  31 & 29\\
Dining      & 166 &  40 & 39\\
Science     & 151 &  68 & 64\\
Upshot      & 146 & 124 & 115\\
Well        & 142 &  47 & 45\\
Metro       & 137 &  53 & 48\\
Business    & 109 & 269 & 239\\
Insider     & 105 &  28 & 27\\
\hline
\end{tabular}
\caption{\textbf{Categories with at least 100 articles in the BBC data (top) and the NYT data (bottom).} Article categories are provided directly by both media outlets.}
\label{tab:categories}
\end{table*}

\begin{table*}
\centering
\begin{tabular}{rrrp{275pt}}
\hline
Rank & Impact & Category & News title\\
\hline
 1 & 3538 & football & Jose Mourinho: Manchester United sack manager\\
 2 & 2241 & football & Liverpool 4-0 Barcelona (4-3 agg): Jurgen Klopp's side complete extraordinary comeback\\
 3 & 1931 & football & Chris Hughton: Brighton sack manager after 17th-placed finish in Premier League\\
 4 & 1874 & football & Ajax 2-3 Tottenham (3-3 on aggregate - Spurs win on away goals): Lucas Moura scores dramatic winner\\
 5 & 1697 & football & Champions League: PSG 1-3 Man Utd (agg: 3-3)\\
 6 & 1659 & football & Manchester City 4-3 Tottenham Hotspur (4-4 agg): Spurs stun City on away goals in modern classic\\
 7 & 1575 & football & Liverpool, Tottenham, Chelsea and Arsenal fans criticise Uefa for final ticket numbers\\
 8 & 1417 & football & Ole Gunnar Solskjaer was the wrong choice as Man Utd manager - Jenas\\
 9 & 1342 &   tennis & Andy Murray: Australian Open could be last tournament\\
10 & 1291 & football & Liverpool beat Spurs 2-0 to win Champions League final in Madrid\\
\hline
 1 & 3983 & national & Trump Intensifies Criticism of F.B.I. and Journalists\\
 2 & 3277 & national & Trump Fires Comey Amid Russia Inquiry\\
 3 & 3077 & business & Man Is Dragged From a Full Jet, Stirring a Furor\\
 4 & 2731 & national & G.O.P. Revolt Sinks Bid to Void Health Law\\
 5 & 2692 & national & Judge Blocks Trump Order On Refugees\\
 6 & 2618 & foreign & Trump Is Said to Expose Ally's Secrets to Russians\\
 7 & 2572 & national & Britain Furious as Trump Pushes Claim of Spying\\
 8 & 2388 & national & Trump Was Told of Claims Russia Has Damaging Details on Him\\
 9 & 2327 & national & Trump Appealed to Comey to Halt Inquiry Into Aide\\
10 & 2304 & national & Trump Fires Justice Chief Who Defied Him\\
\hline
\end{tabular}
\caption{\textbf{Discussions with the highest impact as measured by the number of unique commenting users.} The BBC data (top) and the NYT data (bottom).}
\label{tab:BBC_top}
\end{table*}

\begin{figure}
\centering
BBC:\\[4pt]
\includegraphics[width = \textwidth]{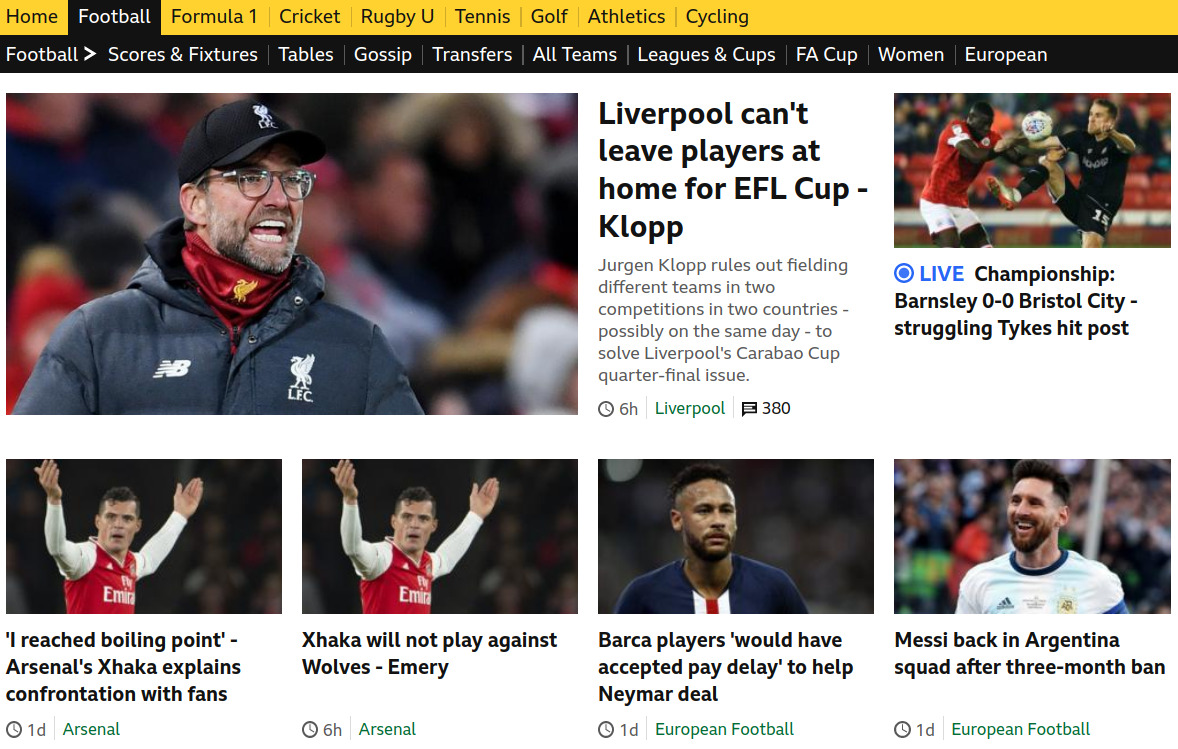}\\[12pt]
NYT:\\[4pt]
\includegraphics[width = \textwidth]{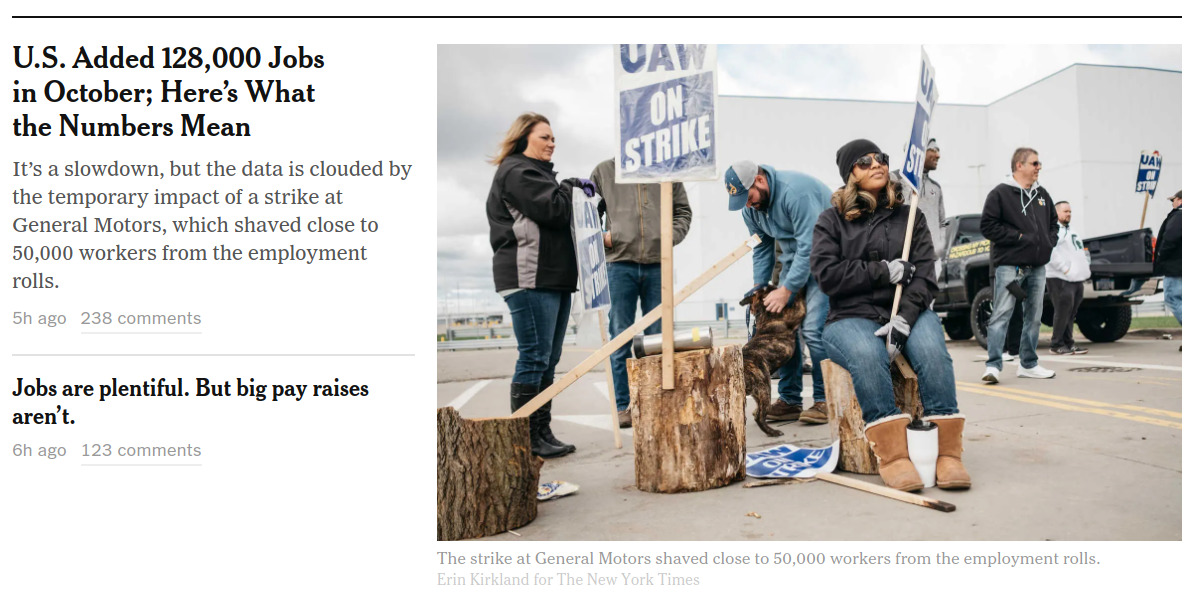}
\caption{\textbf{Snapshots of the BBC website (top) and the NYT website (bottom).} Note that for articles with discussions, the number of comments is indicated for both BBC (icon with number 380 next to it; the other displayed articles do not have discussions) and NYT (labels ``238 comments'' and ``123 comments'').}
\label{fig:snapshots}
\end{figure}

\clearpage
\section{Fitting exponential distributions to article impact data}
The primary fitting results for the BBC data are shown in Table~\ref{tab:fit_exp_BBC} where article impact is measured by the number of unique users who comment on a news article. Since a minority of comments are discarded in this way, the fitting results are qualitatively similar when the total number of comments is used instead. Our fitting procedure follows the steps described in \cite{clauset2009power}: We choose the lower bound, $\cmin$, that minimizes the standard Kolmogorov-Smirnov statistic. The scaling parameter $\lambda$ is then obtained by maximizing the data likelihood for the exponential model. The standard error of this estimate is then obtained by taking the standard deviation of scaling parameters estimated in bootstrap samples of the original data. Finally, the $p$-values characterizing the goodness-of-fit are obtained by comparing the corresponding Kolmogorov-Smirnov statistic measured on the real data with the Kolmogorov-Smirnov statistic measured on data drawn from the exponential distribution with the previously determined lower bound $\cmin$ (which directly influences the sample size represented by the number of article discussions that match or exceed $\cmin$) and the scaling parameter $\lambda$. Upon generating a large number of exponentially distributed samples, the $p$-value is the fraction of the samples that have a higher Kolmogorov-Smirnov statistic than the value found in the real data. A low $p$-value is thus an indication that the artificial exponentially-distributed samples match the fitted exponential distribution better then the real data. $p$-values above 0.10 are conventionally understood as an indication that the fit is good. Needless to say, the $p$-values are jointly influenced by the quality of fit and the sample size. As the sample size grows, the same deviation from the exponential distribution results in progressively lower $p$-values.

\begin{table}[h!]
\centering
{\small\begin{tabular}{lrrrrrrr}
\hline
Category &   $N$ & $O$ & $\cmin$ & $\fmin$ & $KS$ & $\lambda$ & $p$-value\\
\hline
Boxing      &   103 & 0 &   6 & 0.98 & 0.041 & $149\pm15$ & 0.97\\
Cricket     &   240 & 0 &  38 & 0.90 & 0.043 & $107\pm8$  & 0.58\\
Football    & 1,590 & 1 & 381 & 0.18 & 0.026 & $270\pm17$ & 0.96\\
Formula 1   &   139 & 0 &  42 & 0.94 & 0.052 & $180\pm16$ & 0.69\\
Golf        &   123 & 0 &   1 & 1.00 & 0.082 &  $84\pm10$ & 0.13\\
Rugby-union &   439 & 0 &  13 & 0.97 & 0.033 & $121\pm7$  & 0.49\\
Tennis      &   162 & 1 &  97 & 0.32 & 0.044 & $124\pm17$ & 1.00\\
\hline
\end{tabular}}
\caption{\textbf{Results of fitting exponential distributions to individual article categories: BBC data, article impact measured by the number of unique commenting users.} The displayed characteristics are: the number of articles with a discussion, $N$, the number of outliers, $O$, the determined lower bound of the exponential tail, $\cmin$, the fraction of articles that comprise the exponential tail, $\fmin$, the lowest Kolmogorov-Smirnov statistic, $KS$, the determined scaling parameter, $\lambda$, together with its standard error, and the $p$-value of the fit ($p$-values below some threshold, such as 0.10, would indicate that the exponential fit of the corresponding data is of low quality; for $p$-values above the threshold, the exponential distribution is not ruled out).}
\label{tab:fit_exp_BBC}
\end{table}

A single outlier---a news article with 3538 unique users contributing to the discussion---has been identified for the football category in the BBC data. When this news is included in the fitting procedure described above, we obtain $\cmin=7$ and $\hat\lambda=231$. The probability that the value of at least 3538 is observed in a single draw from the exponential distribution with these parameters is $p_0\approx 2.3\cdot 10^{-7}$. With 1577 independent draws from the distribution, corresponding to 1577 news that pass the impact threshold $\cmin=7$, the probability that at least one of them has impact 3538 or more is $1 - (1-p_0)^{1577}\approx 3.7\cdot 10^{-3}$. In other words, this single observation is very unlikely given the fitted exponential distribution which indicates that it is an outlier. The situation is similar in the tennis category where a single news with impact 1342 has probability $8.4\cdot 10^{-3}$ to emerge for the fitted parameters $\cmin=92$ and $\hat\lambda=142$ (unlike Table~\ref{tab:fit_exp_BBC}, these results correspond to fitting all data including the outlier). In all other categories, the largest impact value has probability more than $0.01$ to be observed.

Interestingly, the two identified outliers transcendent the sport boundaries which could help them reach the outstanding impact. The football category outlier is the news ``Jose Mourinho: Manchester United sack manager'' concerning a successful yet controversial coach and the tennis category outlier is the news ``Andy Murray: Australian Open could be last tournament'' concerning the potential end of career of the British most successful tennis player in the modern history.

\begin{figure}[h!]
\centering
\includegraphics[scale=0.7]{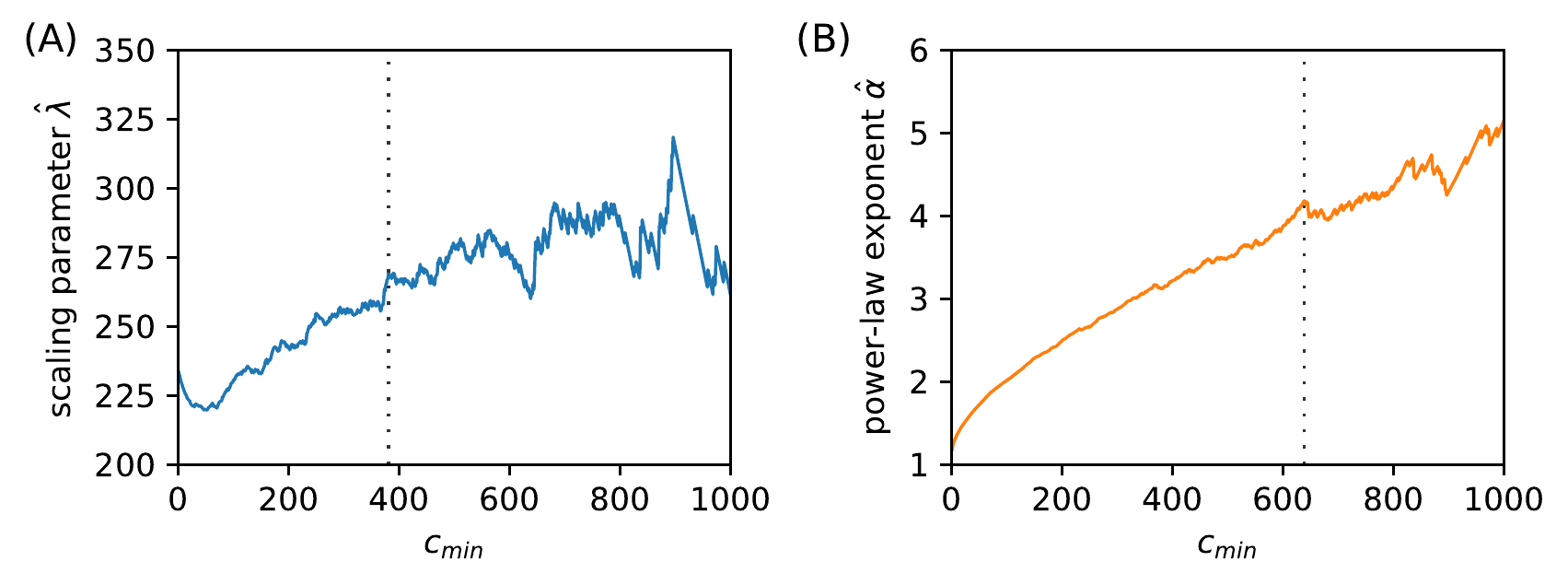}
\caption{\textbf{The dependence of the parameter estimates on the fitting lower bound: a comparison between the exponential and power-law distribution, football category in the BBC data.} The exponential scaling parameter MLE estimate, $\hat\lambda$, varies little over a broad range of the fitting lower bounds, $\cmin$. By contrast, the power-law exponent MLE estimate, $\hat\alpha$, grows steadily and substantially with $\cmin$ which indicates that a power-law is not a good fit of the data. In both panels, the vertical dashed line shows the lower bound estimates which are obtained by minimizing the Kolmogorov-Smirnov statistic with respect to $\cmin$.}
\label{fig:S1_xmin_dependence_BBC}
\end{figure}

The same fitting analysis using power-law distributions shows that they yield worse fit of the data than exponential distributions. This can be visually appreciated in Figure~\ref{fig:S1_xmin_dependence_BBC} which shows the dependence of estimated parameters on the fitting lower bound, $\cmin$. While the estimated scaling parameter of the exponential distribution, $\hat\lambda$, varies in the narrow range $[225, 290]$, estimated exponent of the power-law distribution, $\hat\alpha$, continually grows with $\cmin$ from an extremely low value of 1.2 to more than five and shows no plateau. The higher stability of $\hat\lambda$ as compared to $\hat\alpha$ is a sign that the exponential fits are more robust (less sensitive to the choice of $\cmin$, in particular). Finally, the smallest Kolmogorov-Smirnov statistic for a power-law fit of the most populated football category news yields $\cmin=638$, so the power-law spans over less than one order of magnitude unlike other broadly distributed datasets~\cite{clauset2009power}.

\begin{table}[h!]
\centering
\begin{tabular}{lrrrrrr}
\hline
& \multicolumn{4}{c}{Power-law fit} & \multicolumn{2}{c}{Likelihood ratio test}\\
Category    &  $\cmin$ & $\fmin$ & $KS$ & $\hat\alpha$ & $LR$ & $p$-value\\
\hline
Boxing      & 233 & 0.24 & 0.054 & 3.75 &  13.5 & $0$\\
Cricket     & 147 & 0.32 & 0.057 & 3.11 &  38.2 & $0$\\
Football    & 638 & 0.07 & 0.068 & 4.18 &  64.2 & $0$\\
Formula1    & 141 & 0.58 & 0.081 & 2.54 &  29.4 & $0$\\
Golf        &  83 & 0.33 & 0.059 & 2.64 &  11.3 & $0$\\
Rugby union & 279 & 0.12 & 0.065 & 4.03 &  36.4 & $0$\\
Tennis      & 189 & 0.17 & 0.076 & 3.15 &  14.1 & $0$\\
\hline
\end{tabular}
\caption{\textbf{Results of fitting power-law distributions to individual article categories and the likelihood ratio tests, BBC data.} We measure article impact by the number of unique users who have commented on it and use the standard Kolmogorov-Smirnov statistic for the fitting analysis. We show here the results of the power-law fitting (the estimated lower bound, the fraction of article discussions that comprise the identified power-law tail, and the estimated power-law exponent) as well as the likelihood test results (the normalized log likelihood ratio, $LR$, that measures the difference in how well the fits agree with the data and the corresponding $p$-values that estimates how likely it is to find $LR$ as high or higher by chance).}
\label{tab:fit_pwl_BBC}
\end{table}

Table~\ref{tab:fit_pwl_BBC} further reports detailed results of fitting with power-law distributions. We see that the lower bounds obtained for power-law fits are substantially higher than the corresponding values for exponential fits, which shows that exponential distributions fit greater portions of respective datasets. The Kolmogorov-Smirnov statistic values obtained with exponential fits are lower than those obtained with power-law fits, indicating better match between the data and the fitted distributions. We conclude the assessment of power-law and exponential fits by a direct comparison between the exponential and power-law distribution in terms of how well they fit the data (to allow for a fair comparison, we do not exclude any outliers here as that would put the power-law hypothesis in a disadvantage). This can be done using the log-likelihood test described in~\cite{clauset2009power}. As can be seen in Table~, the normalized log likelihood ratio~\cite{vuong1989likelihood}, $LR$, favors the exponential distribution for all analyzed news categories and the corresponding $p$-values show that the obtained $LR$ values are highly significant ($p$-values $10^{-28}$ and less).

\begin{figure}[h!]
\centering
\includegraphics[width=\textwidth]{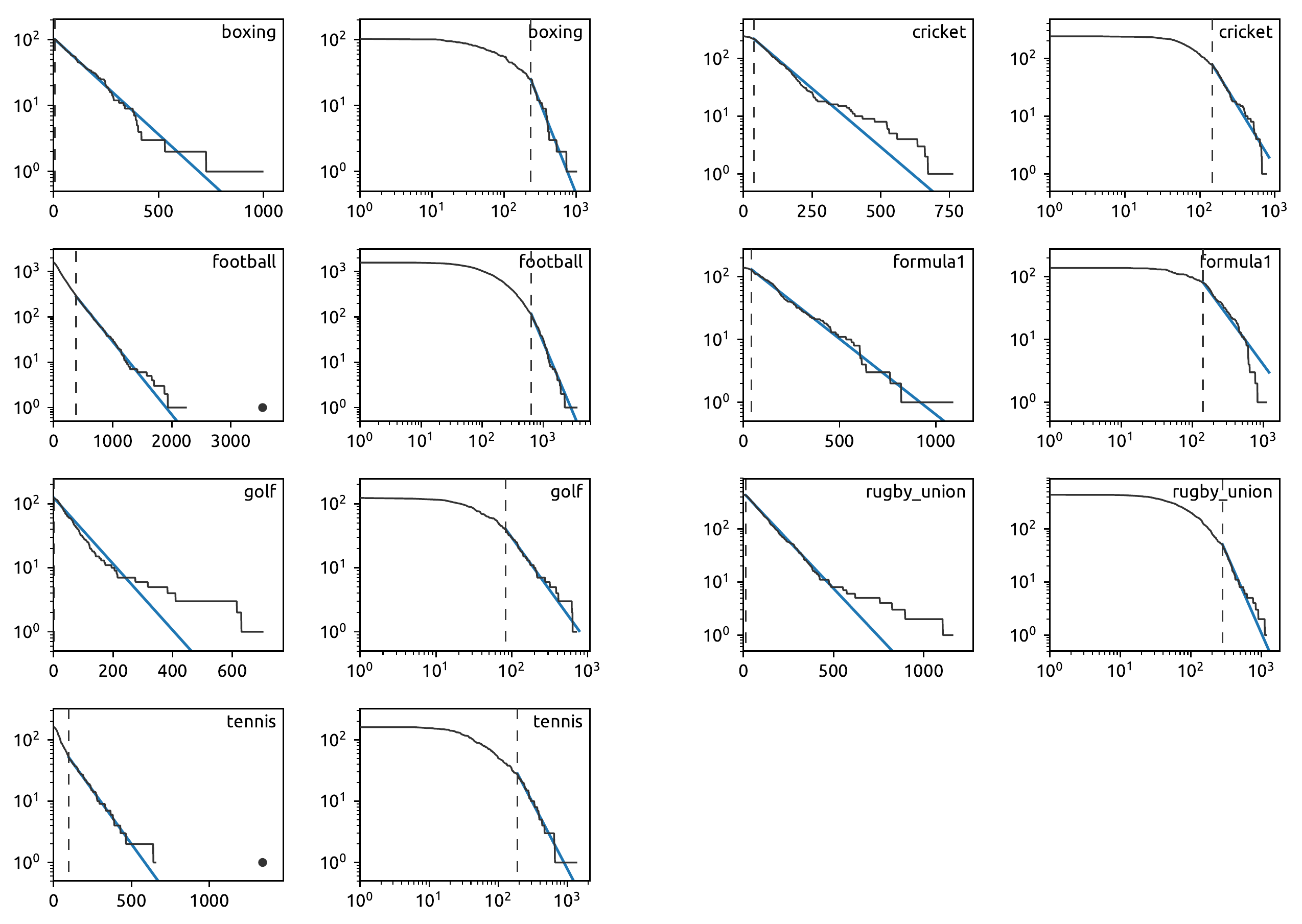}
\caption{\textbf{Visualizations of MLE fits for individual news categories, BBC data.} Each news category is displayed twice: in the log-linear scale (left) with the exponential fit and in the log-log scale (right) with the power-law fit. The vertical dashed lines mark the estimated lower fitting bounds.}
\label{fig:all_categories_BBC}
\end{figure}

Finally, Figure~\ref{fig:all_categories_BBC} visually compares exponential and power-law fit for all news categories. While two categories---Golf and Rugby Union---display an excess of high-impact news, those high-impact news are fewer than ten in both cases, so they do not suffice to turn around the likelihood ratio test. The piece-wise linear shape of the distributions for Football and Tennis suggests that their fits could be improved by using a combination of two exponentials with different scale parameters, corresponding to two different classes of news mixed in the same category. Interestingly, the same pattern can be seen also for some NYT news categories in Figure~\ref{fig:all_categories_NYT} (categories Dining and Foreign, for example). This suggests the hypothesis that while a homogeneous news category is well fitted with a single exponential, a mixture of two exponential distributions may be more appropriate for a heterogeneous news category. Further work is needed to explore such possible fine structure of news categories.

\begin{table}[h!]
\centering
{\small\begin{tabular}{lrrrrrrr}
\hline
Category &   $N$ & $O$ & $\cmin$ & $\fmin$ & $KS$ & $\lambda$ & $p$-value\\
\hline
Business &  109 &   1 &   0 & 0.99 & 0.064 & $213\pm23$  & 0.52\\
Dining   &  166 &   0 &  31 & 0.36 & 0.068 &  $54\pm6$   & 0.80\\
Foreign  &  204 &   0 & 214 & 0.31 & 0.060 & $396\pm52$  & 0.91\\
Games    &  192 &   0 &  35 & 0.34 & 0.056 & $6.4\pm0.9$ & 0.95\\
Insider  &  105 &   0 &   2 & 0.97 & 0.081 & $26\pm3$    & 0.22\\
Learning &  306 &   0 &  72 & 0.09 & 0.148 & $352\pm72$  & 0.27\\
Magazine &  262 &   0 & 118 & 0.13 & 0.071 & $217\pm35$  & 0.97\\
Metro    &  137 &   0 &  47 & 0.26 & 0.083 & $96\pm17$   & 0.86\\
National &  348 &   0 &  28 & 0.97 & 0.037 & $567\pm32$  & 0.53\\
Science  &  151 &   1 &   0 & 0.99 & 0.049 & $60\pm4$    & 0.63\\
Sports   &  213 &   0 &  49 & 0.25 & 0.081 & $78\pm12$   & 0.65\\
Upshot   &  146 &   1 &  34 & 0.84 & 0.062 & $90\pm9$    & 0.47\\
Well     &  142 &   0 &   3 & 0.99 & 0.050 & $43\pm3$    & 0.65\\
\hline
\end{tabular}}
\caption{\textbf{Results of fitting exponential distributions to article categories: NYT data, article impact measured by the number of unique users at the top commenting level.} Notation as in Table~\ref{tab:fit_exp_BBC}.}
\label{tab:fit_exp_nyt}
\end{table}

\begin{figure}[h!]
\centering
\includegraphics[scale=0.7]{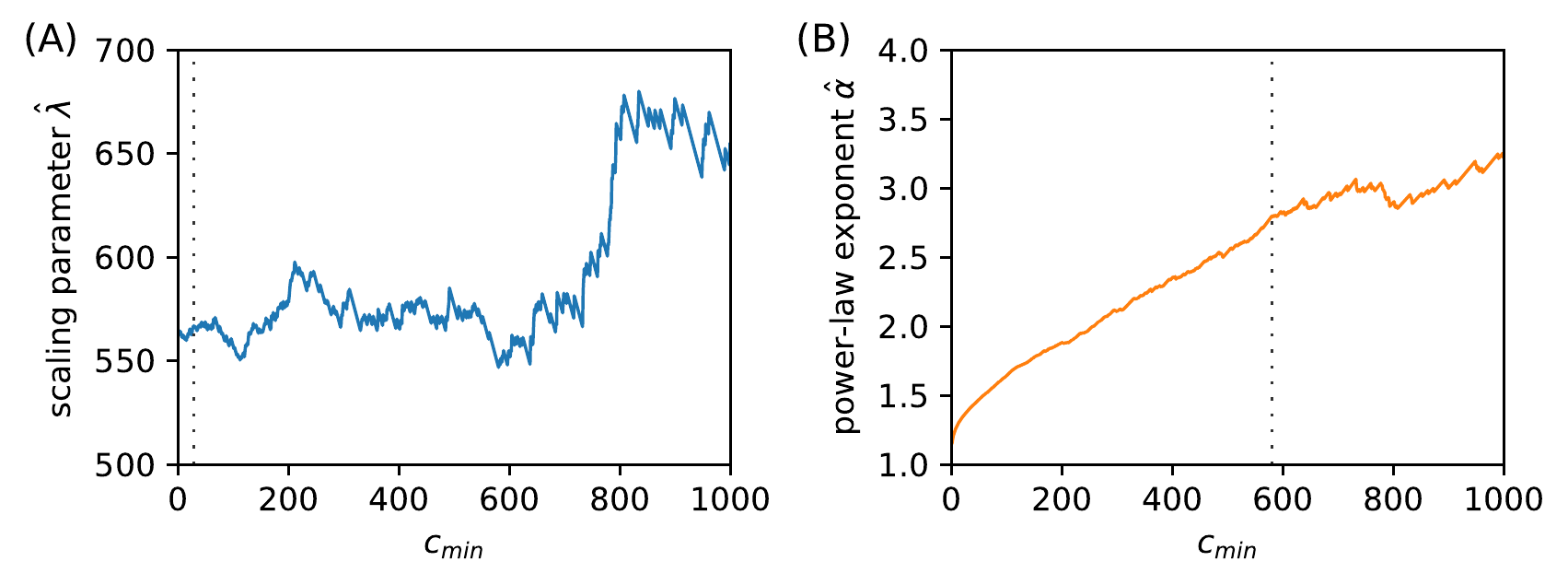}
\caption{\textbf{The dependence of the parameter estimates on the fitting lower bound: a comparison between the exponential and power-law distribution, National category in the NYT data.} The exponential scaling parameter MLE estimate, $\hat\lambda$, varies little over a broad range of the fitting lower bounds, $\cmin$. By contrast, the power-law exponent MLE estimate, $\hat\alpha$, grows steadily and substantially with $\cmin$ which indicates that a power-law is not a good fit of the data. In both panels, the vertical dashed line shows the lower bound estimates which are obtained by minimizing the Kolmogorov-Smirnov statistic with respect to $\cmin$.}
\label{fig:S2_xmin_dependence_NYT}
\end{figure}

\begin{figure}[h!]
\centering
\includegraphics[width=\textwidth]{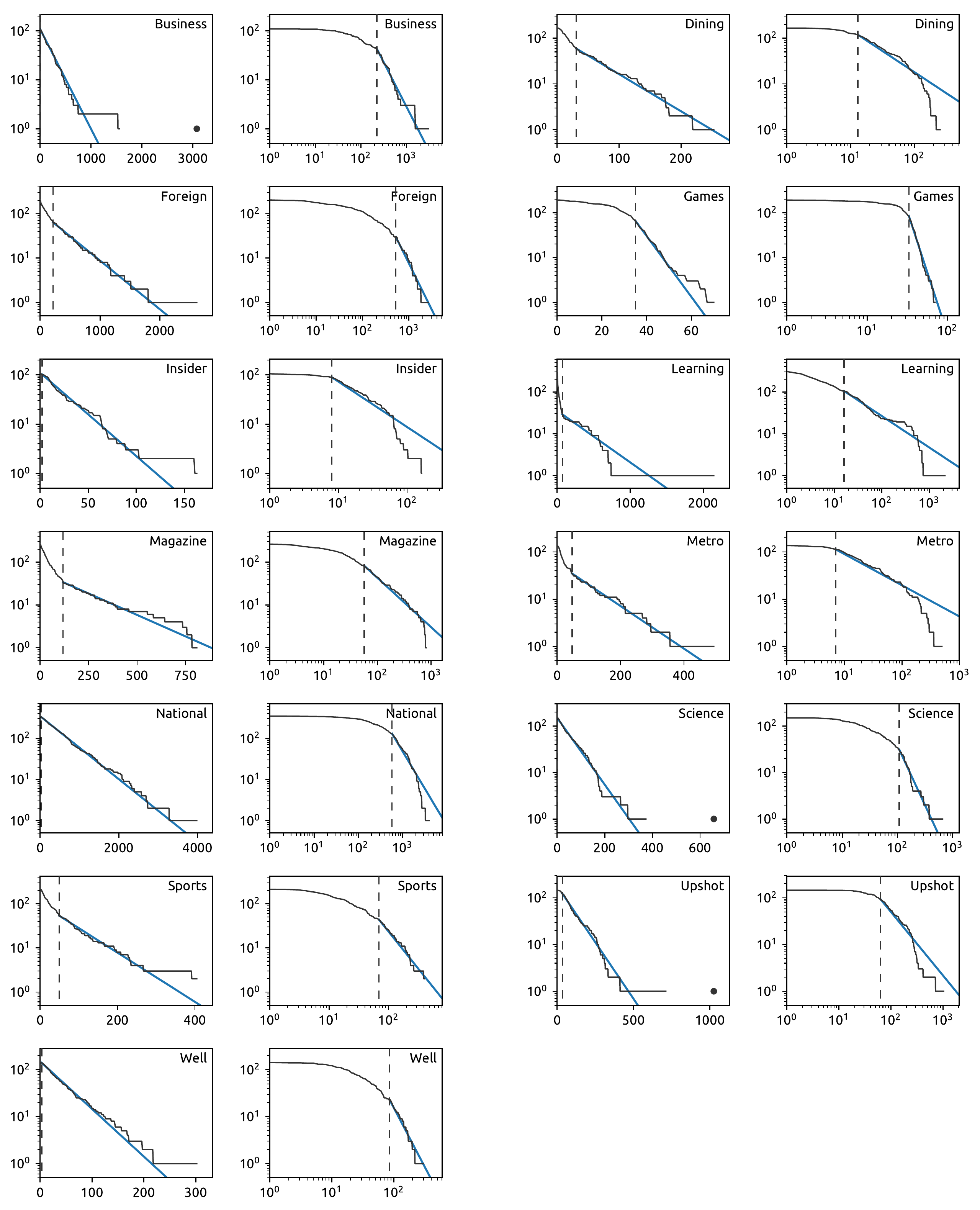}
\caption{\textbf{Visualizations of MLE fits for individual news categories, NYT data.} Each news category is displayed twice: in the log-linear scale (left) with the exponential fit and in the log-log scale (right) with the power-law fit. The vertical dashed lines mark the estimated lower fitting bounds.}
\label{fig:all_categories_NYT}
\end{figure}

We applied the same steps to the NYT data where the article impact is measured by the number of unique commenting users at the top commenting level (see Section~\ref{sec:data_description} for details), focusing on the 13 news categories with at least 100 news. Table~\ref{tab:fit_exp_nyt} summarizes the results of exponential fitting, showing that the exponential distribution again cannot be ruled out for any news category, albeit now there are two categories---Learning and Magazine---where only a small fraction of news belong to the exponential tail and the best Kolmogorov-Smirnov statistic for Learning news is high (0.148). Figure~\ref{fig:S2_xmin_dependence_NYT} illustrates the stability of the scaling parameter estimate over a broad range of the lower fitting bound, $\cmin$, whereas the estimated power-law exponents substantially grows with $\cmin$. Finally, Table~\ref{tab:fit_pwl_NYT} summarizes results of power-law fits and likelihood ratio tests for individual article categories. Category Learning is confirmed here as the only one for which the exponential fit is worse than the power-law fit (normalized log likelihood ratio, $LR$, is significantly negative; see Sec.~\ref{sec:learning} for more information on this category). For the remaining 12 categories, exponential fits are much more apt than power-law fits. Figure~\ref{fig:all_categories_NYT} visually compares exponential and power-law fit for all news categories.

\begin{table}[h!]
\centering
\begin{tabular}{lrrrrrr}
\hline
& \multicolumn{4}{c}{Power-law fit} & \multicolumn{2}{c}{Likelihood ratio test}\\
Category    &  $\cmin$ & $\fmin$ & $KS$ & $\hat\alpha$ & $LR$ & $p$-value\\
\hline
Business   & 224 & 0.39 & 0.086 & 2.82 &  9.7 & 0\\
Dining     &  13 & 0.71 & 0.095 & 1.92 & 13.2 & 0\\
Foreign    & 521 & 0.15 & 0.100 & 3.11 &  6.2 & 0\\
Games      &  33 & 0.44 & 0.063 & 6.48 & 31.5 & 0\\
Insider    &   8 & 0.85 & 0.101 & 1.92 & 10.5 & 0\\
Learning   &  16 & 0.34 & 0.074 & 1.75 & $-4.9$ & 0\\
Magazine   &  57 & 0.31 & 0.070 & 2.15 &  4.8 & 0\\
Metro      &   7 & 0.83 & 0.095 & 1.66 &  4.9 & 0\\
National   & 580 & 0.37 & 0.081 & 2.80 & 24.4 & 0\\
Science    & 108 & 0.21 & 0.115 & 3.56 & 12.0 & 0\\
Sports     &  69 & 0.21 & 0.069 & 2.68 &  8.1 & 0\\
Upshot     &  63 & 0.64 & 0.084 & 2.36 & 22.2 & 0\\
Well       &  85 & 0.16 & 0.087 & 3.52 & 14.3 & 0\\
\hline
\end{tabular}
\caption{\textbf{Results of fitting power-law distributions to individual article categories and the likelihood ratio tests, NYT data.} We measure article impact by the number of unique users who have commented on it and use the standard Kolmogorov-Smirnov statistic for the fitting analysis. We show here the results of the power-law fitting (the estimated lower bound, the fraction of article discussions that comprise the identified power-law tail, and the estimated power-law exponent) as well as the likelihood test results (the normalized log likelihood ratio, $LR$, that measures the difference in how well the fits agree with the data and the corresponding $p$-values that estimates how likely it is to find $LR$ as high or higher by chance).}
\label{tab:fit_pwl_NYT}
\end{table}

\subsection{Specifics of the category ``Learning'' in the NYT data}
\label{sec:learning}
The Learning category has a particularly uneven impact distribution: While the median impact over its 306 news is only 7, there are 23 news with impact above 100 and the top impact is 2,344. Nevertheless, it turns out that the Learning category actually does not contain news but a mixed bag of content. Among the 23 news with impact above 100, there are 17 contest articles ``What’s Going On in This Picture?'' (published on a weekly basis) with the median impact of 505 and the top impact article is yet another contest.\footnote{\url{https://www.nytimes.com/2017/04/05/learning/our-eighth-annual-found-poem-student-contest.html}}
By contrast, little impact Learning articles are ``teaching and learning resources based on New York Times journalism'' made available under the name ``Learning Network''.\footnote{See \url{https://www.nytimes.com/2017/04/18/learning/what-does-a-presidents-choice-of-pet-or-choice-not-to-have-a-pet-at-all-say-about-him.html} for an example.}

\clearpage
\section{Elementary analysis of the system's dynamics}
\begin{figure}[h!]
\centering
\includegraphics[scale=0.7]{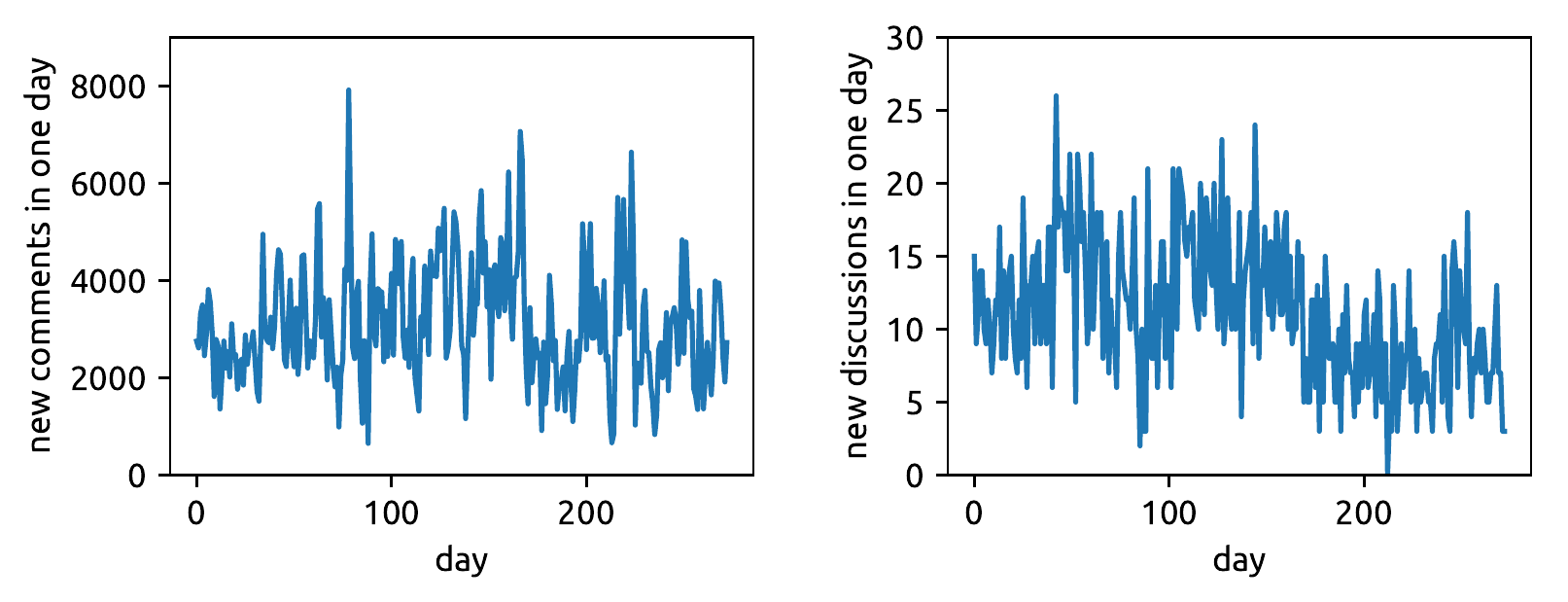}
\includegraphics[scale=0.7]{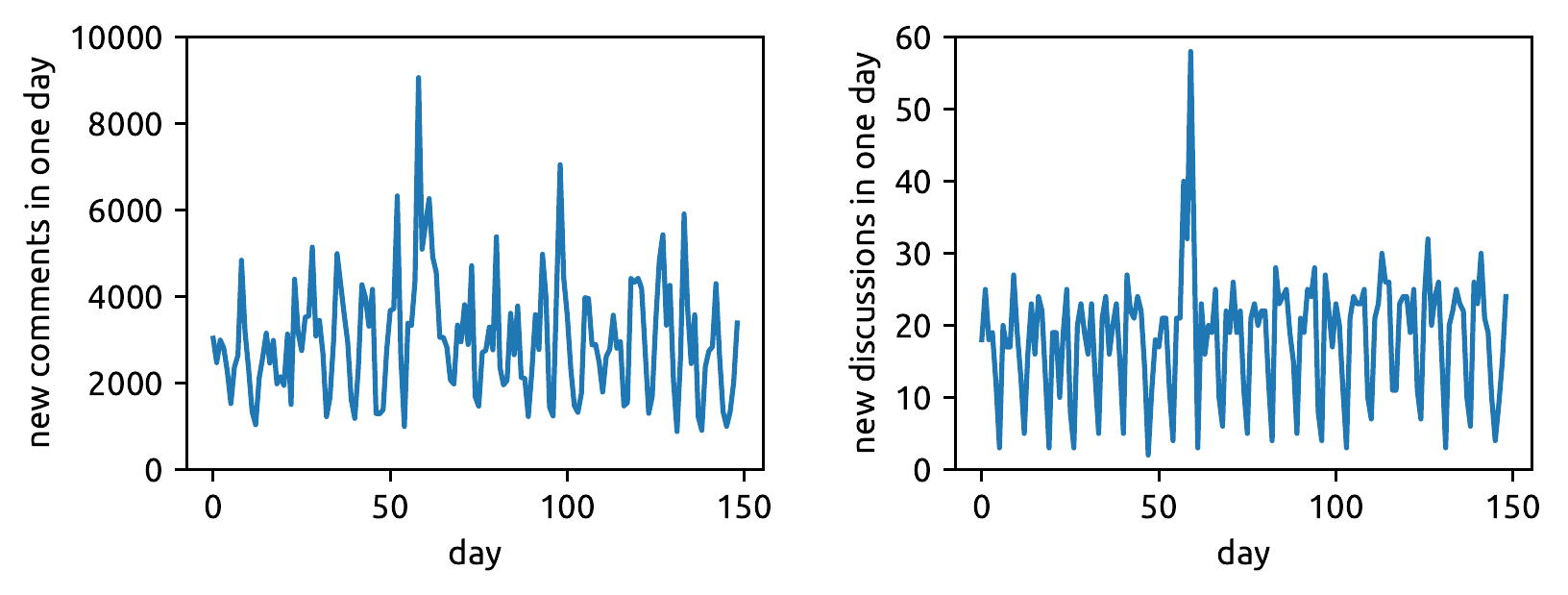}
\caption{\textbf{Variations of the daily commenting activity.}\newline
(Top row, BBC data) There are 3,109 new comments a day on average, the standard deviation is 1,215. There are 11.3 new article discussions a day on average, the standard deviation is 4.8.\newline
(Bottom row, NYT data) There are 3,007 new comments a day on average, the standard deviation is 1,347. There are 18.4 new article discussions a day on average, the standard deviation is 8.2.}
\label{fig:SI_activity}
\end{figure}

\begin{figure}[h!]
\centering
\includegraphics[scale=0.7]{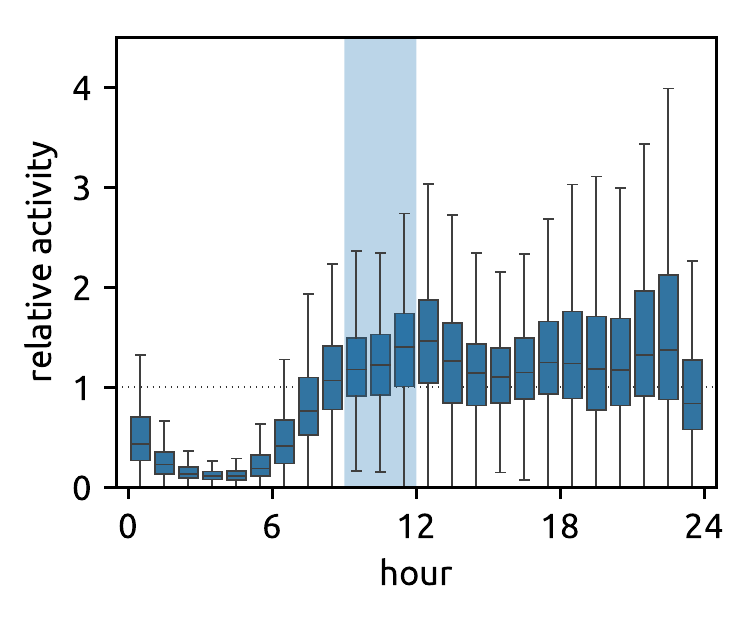}
\includegraphics[scale=0.7]{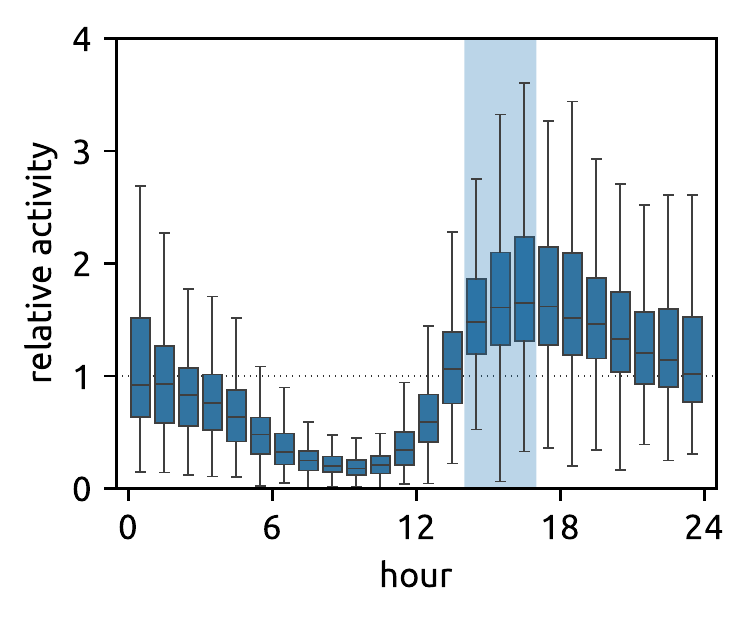}
\caption{\textbf{Variations of the normalized commenting activity during the day.} The value of one corresponds to 1/24 of the day's comments arriving in a given hour.\newline
(Left, BBC data) The activity is low from 1am to 6am and relatively constant between 8am and midnight. The shaded area shows the time of day that we use for the analysis of the commenting dynamics: These articles have a long period of approximately uniform website activity before the night arrives and the activity drops.\newline
(Right, NYT data) Due to the time difference, the commenting activity is lower between 5am and noon. The shaded area again shows the ``morning articles'' that are used for the analysis of new dynamics.}
\label{fig:SI_daily_activity}
\end{figure}

\begin{figure}[h!]
\centering
\includegraphics[scale=0.7]{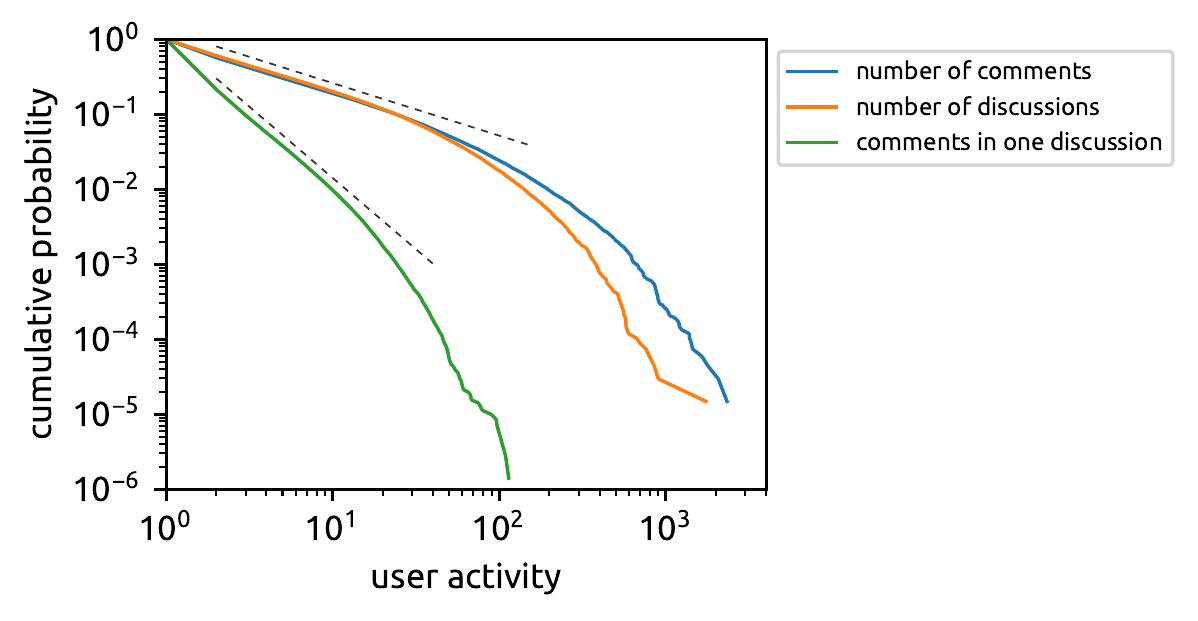}
\includegraphics[scale=0.7]{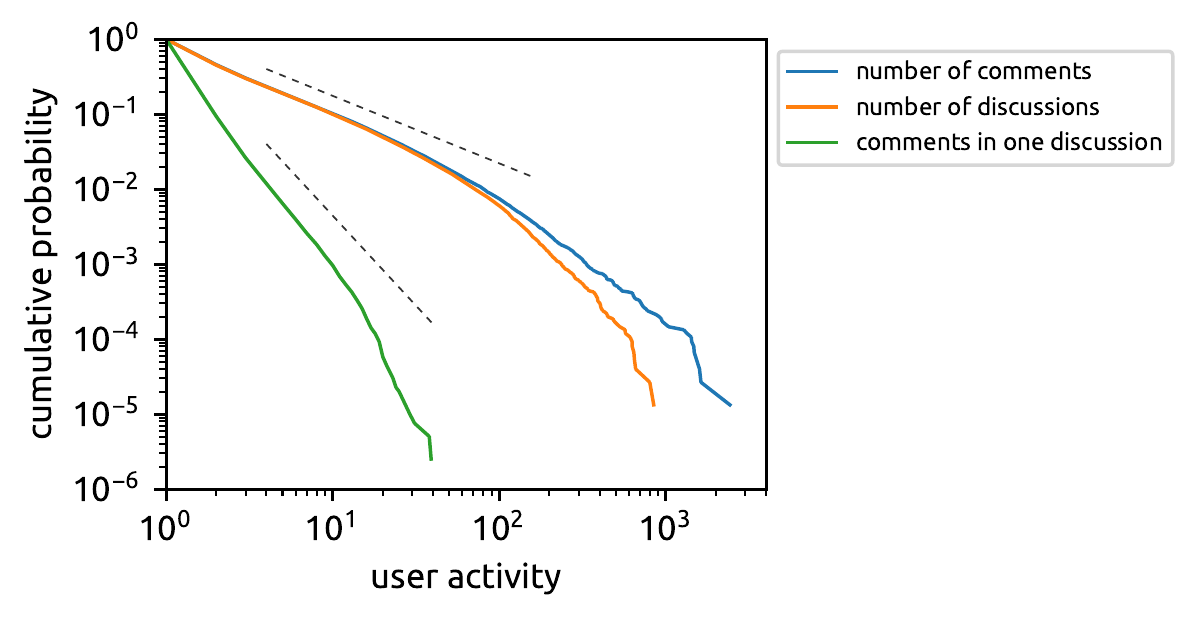}
\caption{\textbf{The distributions of various measures of user activity.}\newline
(Top, BBC data) The indicative dashed lines have the slopes of $0.7$ and $1.9$, respectively.\newline
(Bottom, NYT data) The indicative dashed lines have the slopes of $0.9$ and $2.4$, respectively.\newline
All three distributions have an initial power-law part with a cut-off at higher activity values.}
\label{fig:SI_user_activity}
\end{figure}

We present here basic characteristics of the commenting dynamics in the BBC and the NYT data: the daily and hourly profiles of user activity and the distributions of user activity (Figures~\ref{fig:SI_activity}--\ref{fig:SI_user_activity}).

\clearpage
\section{Preferential attachment in measurements and models}
\label{sec:PA}
\begin{figure}[h!]
\centering
\includegraphics[scale=0.7]{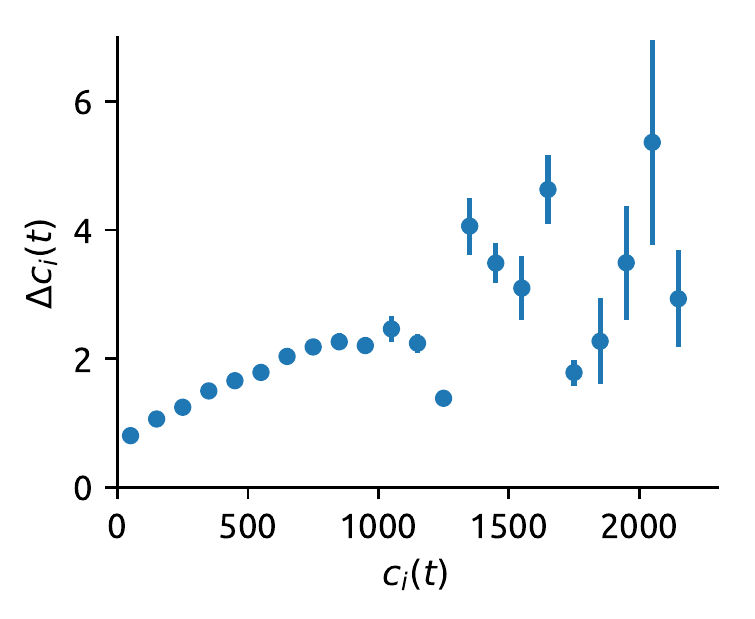}
\caption{\textbf{Measuring preferential attachment in synthetic data without preferential attachment.} We use here the synthetic dataset produced using the model described in Section~\ref{sec:model} without preferential attachment. The fit up to the comment count 800 yields the slowly-growing dependence proportional to $1+c_i(t)/336$. Since there is no preferential attachment, the observed weak dependence of $\Delta c_i(t)$ on $c_i(t)$ is an artifact of the model's dynamics.}
\label{fig:SI_PA_synthetic}
\end{figure}

Figure~2A in the main text shows that preferential attachment is extremely weak in the BBC data with the linear fit for $c_i(t)<800$ in the form $\Delta c_i(t)\sim 1 + c_i(t)/220$ followed by saturation for higher $c_i(t)$ values. We add that a similar weak dependence of $\Delta c_i(t)$ on $c_i(t)$ can be observed in synthetic data where preferential attachment is absent. To demonstrate this, we use the model described in Section~\ref{sec:model} which produces a synthetic dataset whose basic properties are similar to those of the BBC dataset. The measurement of preferential attachmed identical with the one used in the main text yields similar results as those in Figure~3A (see Figure~\ref{fig:SI_PA_synthetic}) and a similar fit $\Delta c_i(t)\sim 1 + c_i(t)/336$. The observed weak growth of $\Delta c_i(t)$ with $c_i(t)$ is thus an artifact of the commenting dynamics produced by the model. In particular, articles with high final degree (comment count) initially increase their $c_i(t)$ fast, so their high $\Delta c_i(t)$ values contribute to the initial low $c_i(t)$ bins for a short time. By contrast, articles with low final degree remain during their whole lifetime in low $c_i(t)$ bins where they contribute with their low $\Delta c_i(t)$ values and hence cause the average $\Delta c_i(t)$ values to be lower in low $c_i(t)$ bins that they are in high $c_i(t)$ bins.

\begin{figure}[h!]
\centering
\includegraphics[scale=0.7]{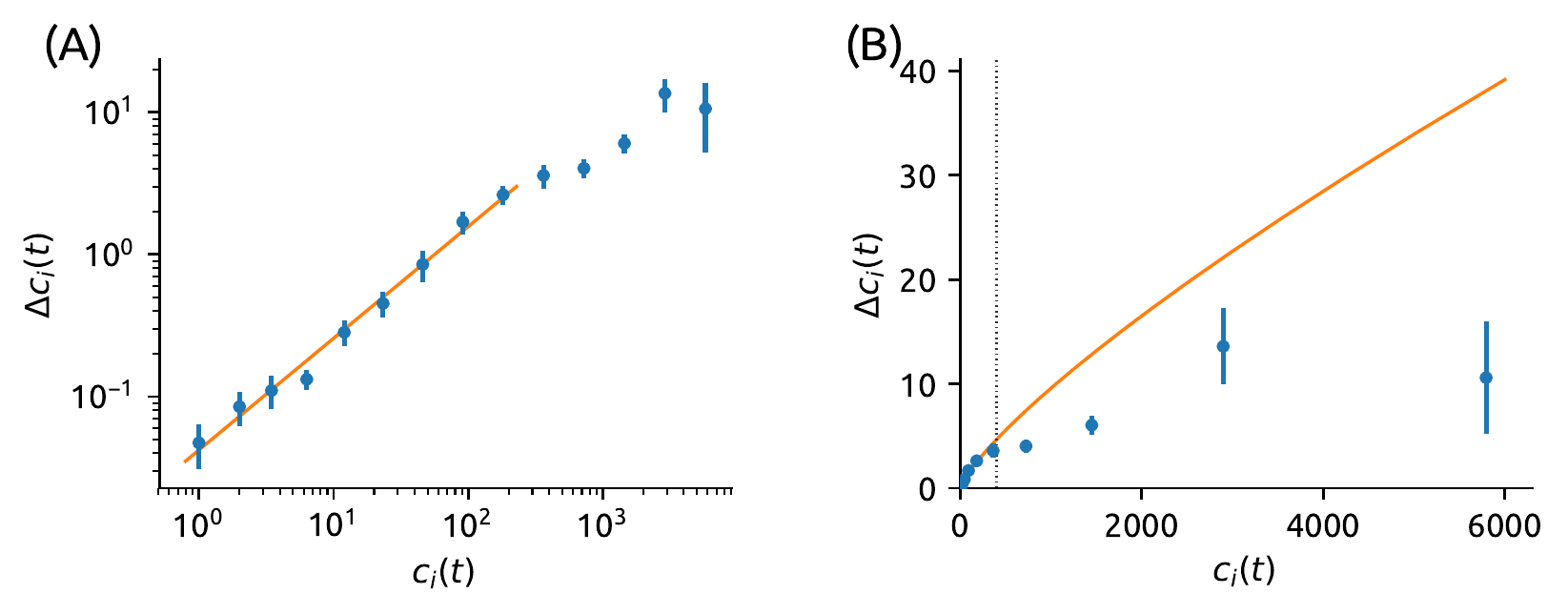}
\caption{\textbf{Non-linear preferential attachment and its saturation in the NYT data.} To study non-linear preferential attachment, we use logarithmic bins on the $x$-axis and log-log scale to plot the same measurement of preferential attachment as in Figure~3B in the main text. (A) The linear fit of the first nine bins (the last bin includes degree values up to 256) has the slope of $0.79$. (B) As panel (A) but with linear axes. The saturation of preferential attachment at, approximately, degree 400 (marked with the vertical dotted line) is well visible here.}
\label{fig:SI_PA_NYT}
\end{figure}

\begin{figure}[h!]
\centering
\includegraphics[scale=0.7]{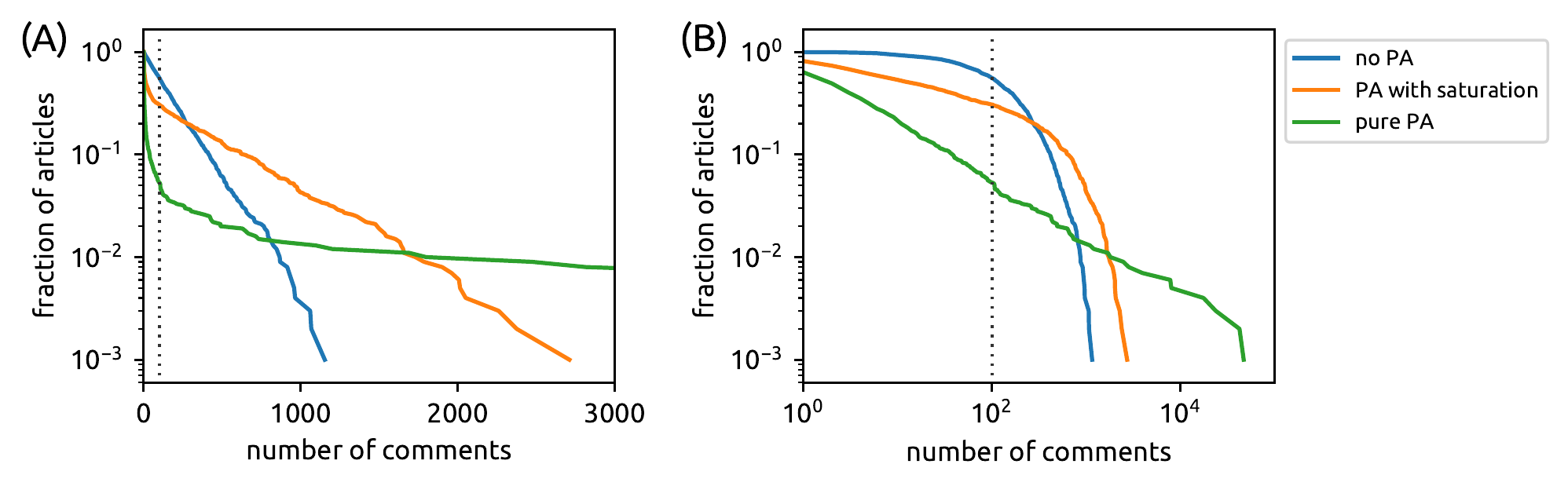}
\caption{\textbf{Degree distributions for various variants of preferential attachment.} The two panels use a log-linear scale (A) and a log-log scale (B), respectively. Degree distributions of networks where the rates of node degree increase are $a \eta_i F(c_i) \exp(-t/\varTheta)$, $\varTheta=240\,\text{min}$. Node fitness is drawn from the exponential distribution, $\rho(\eta) = \exp(-\eta/\lambda)/\lambda$ for $\eta\in[0,\infty)$. We compare no preferential attachment [$F(c)=1$, $\lambda=0.75$], standard preferential attachment [$F(c)=1+c$, $\lambda=0.007$], and preferential attachment with saturation [$F(c)=1+c$ for $c<100$ and $F(c)=100$ for $c\geq 100$, $\lambda=0.018$]. The proportionality rates $a$ are chosen so that the average degree is similar, approximately 180, in all three model variants. The vertical dotted lines mark the saturation threshold $c=100$.}
\label{fig:SI_PA_vs_noPA}
\end{figure}

We claim in the main text that preferential attachment with saturation fails to produce a power-law degree distribution from exponentially distributed fitness values. This can be illustrated using the synthetic data model from Section~\ref{sec:model} where the rate at which an article receives new comments (number of new comments per minute) at time $t$ is assumed to be
\begin{equation}
\label{rate}
\eta_i F[c_i(t)] D(t)
\end{equation}
where $\eta_i$ is the fitness of article $i$, $F[c_i(t)]$ is the degree-dependence term depending on the current number of comments, $c_i(t)$, and $D(t)$ is an aging term. We assume that article fitness is exponentially distributed, $\rho(\eta) = \exp(-\eta/\lambda)/\lambda$ for $\eta\in[0,\infty)$. The distribution's scale, $\lambda$, determines the resulting average comment count. Motivated by Figure~4 in the main text, we assume exponential aging, $D(t) = \exp[-(t-t_i)/\varTheta]$ where $t_i$ is the appearance time of article $i$ and $\varTheta$ is the aging timescale. For the degree-dependence term, we use three different choices. Firstly, degree-independent growth given by $F(c) = 1$. As shown in Sec.~\ref{sec:model} and confirmed in Figure~\ref{fig:SI_PA_vs_noPA}, the degree distribution is then exponential. Secondly, standard preferential attachment given by $F(c) = 1 + c$ (the addition of one is necessary as the initial comment count of all articles is zero). As shown in~\cite{medo2011temporal} and confirmed by the results in Figure~\ref{fig:SI_PA_vs_noPA}, exponentially distributed fitness combined with preferential attachment produces a power-law degree distribution. Finally, preferential attachment with saturation given by $F(c) = 1 + c$ for $c<100$ and $F(c) = 100$ for $c\geq 100$. As shown in Figure~\ref{fig:SI_PA_vs_noPA}, the degree distribution then appears to be a power law (straight in the log-log scale) until $c\approx 100$ and then follows a clear exponential course for larger $c$. The importance of the saturation threshold is natural as the emergence of a power-law degree distribution is a direct consequence of the interplay between the exponential fitness distribution and preferential attachment. Since preferential attachment is absent above the saturation threshold, an exponential degree distribution emerges for large comment counts in the same way as it emerges for all comment counts when preferential attachment is absent entirely [\ie, when $F(c)=1$].

\clearpage
\section{Modeling the dynamics of news impact}
\label{sec:model}
\begin{figure}[h!]
\centering
\includegraphics[scale=0.7]{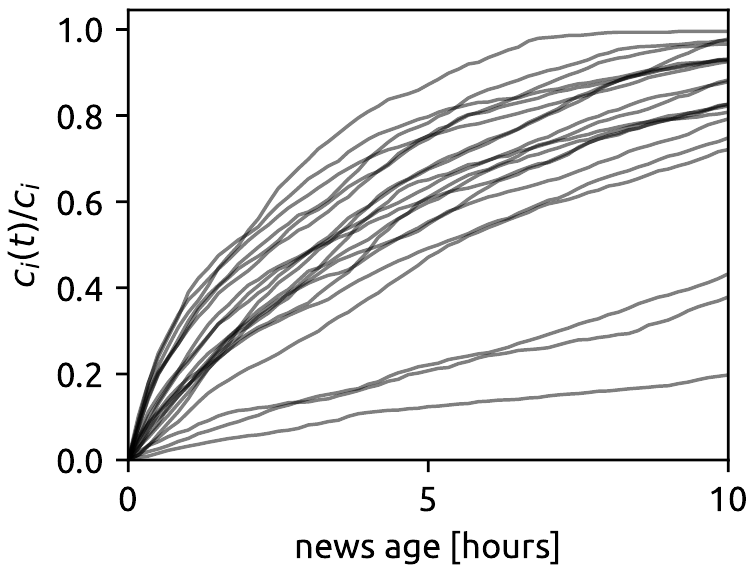}
\caption{\textbf{The degree trajectories of the 20 highest impact articles in the BBC data.} To suppress the time-of-the-day effects during the displayed 10-hour period, we select only the articles that appear in the morning (as in Figure~4 and 5 in the main text). Changing an article's position at the website has the potential to influence its exposure to the users and, in turn, the rate at which the article receives new comments. The displayed individual trajectories do not exhibit substantial changes of the commenting rates which indicates that the effect of changing the articles' positions is minor and can be neglected in the scope of our analysis of commenting dynamics.}
\label{fig:SI_dynamics_top}
\end{figure}

Based on the results presented in the main text, we see that the article commenting dynamics is influenced by two principal factors:
\begin{enumerate}
\item \emph{Absence of preferential attachment,}

\item \emph{Exponential aging,}

\item \emph{Overall user activity at the website.}
\end{enumerate}
We assume for the moment that the overall user activity remains stable during the modeling period. A generalized model which accounts for varying user activity, as discussed in the last section of the main text, is presented in Eq.\,(4) in the main text. We further assume that exogeneous factors such as the article position on the BBC Sport website (on the front page or not, at the top of a page or not, and so forth---see Figure~\ref{fig:snapshots} or \url{https://www.bbc.com/sport} for examples). While article position might seem like an important factor, Figure~\ref{fig:SI_dynamics_top} shows that commenting dynamics of high-impact articles does not display abrupt changes that could correspond to changes of each article's position. We thus neglect article position at the website which, in any case, is not contained in the collected data.

\begin{figure}[h!]
\centering
\includegraphics[scale=0.7]{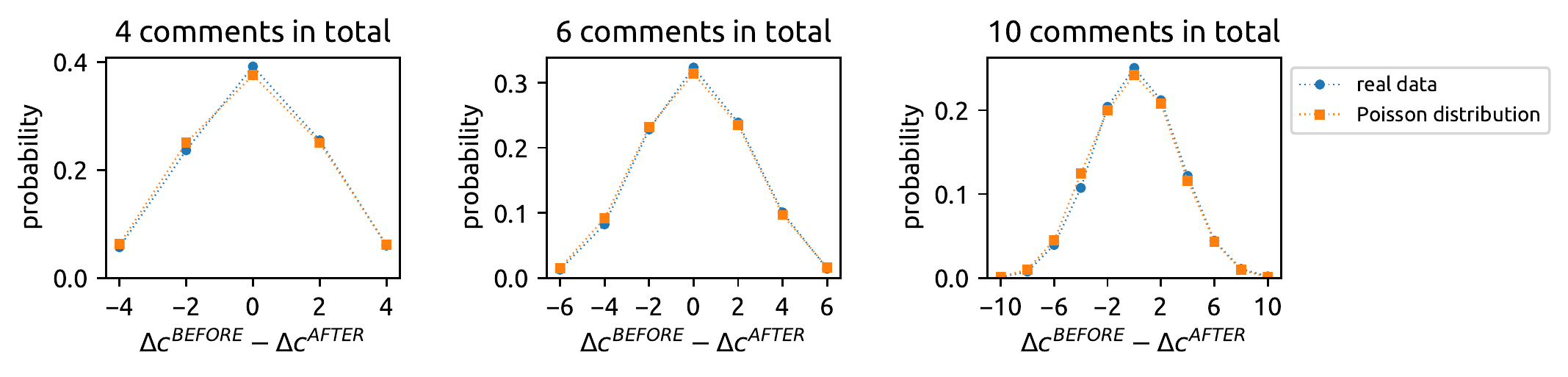}
\caption{\textbf{Probing the comment count randomness in the BBC data.} Using many pairs of consecutive time windows of length $\Delta T=10\,\text{min}$ each, we collected comment count increase values $\Delta c^{BEFORE}$ and $\Delta c^{AFTER}$ for a large number of articles. The histograms compare the empirical distributions of $\Delta c^{BEFORE} - \Delta c^{AFTER}$ with the distributions obtained for Poisson-distributed comment counts. For an easier interpretation of the results, the histograms are plotted separately for observations where $\Delta c^{BEFORE} + \Delta c^{AFTER}$ is 4, 6, and 10, respectively.}
\label{fig:SI_Poisson_verif}
\end{figure}

The factors 1 and 2 in the list above correspond to the expected number of new comments in the form
\begin{equation}
\label{model_expectation}
\overline{\Delta c_i(t)} / \Delta t = \eta_i \exp[-(t-t_i)/\varTheta]
\end{equation}
where $c_i(t)$ is the number of comments of article $i$ at time $t$, $\eta_i$ is the fitness of article $i$ which reflects how attractive it is to the users, $t_i$ is the time when the news article was published (and its discussion opened, $c_i(t_i)=0$), and finally $\varTheta_i$ is the article's aging timescale. The actual number of new comments is drawn from the Poisson distribution with mean given by \eref{model_expectation}. This choice is motivated by independent actions of the users. In Figure~\ref{fig:SI_Poisson_verif}, we illustrate that the Poisson distribution is indeed a good approximation by studying the number of new comments that article discussions receive in two consecutive time windows. The results are grouped by the total number of comments an article discussion receives in the two time windows and we plot the difference of the number of comments between the two time windows. This distribution is then compared with the distribution of differences observed when the comments counts are drawn from a Poisson distribution. As can be seen in the figure, the empirical distributions are indeed similar with the distributions produced by Poisson-distributed comment counts.

The discrete model given by \eref{model_expectation} combined with draws from the Poisson distribution can be replaced with the continuous modeling using the differential equation
\begin{equation}
\label{model_continuous}
\frac{\dd \overline{c_i(t)}}{\dd t} = \eta_i \exp[-(t-t_i)/\varTheta_i]
\end{equation}
with the initial condition $c_i(t_i)=0$. This is referred to as the Fitness-Aging (FA) model in the main text. A similar model was proposed in~\cite{golosovsky2018mechanisms} which explores the relation between fitness and preferential attachment. \eref{model_continuous} is easy to solve and yields
\begin{equation}
\label{average_impact_dynamics}
\overline{c_i(t)} = \eta_i\varTheta_i\biggl(1 - \exp\big[-(t-t_i)/\varTheta_i\big]\biggr).
\end{equation}
with the limit value $\overline{c_i} = \eta_i\varTheta_i$. In the absence of preferential attachment, the final impact of an article is directly proportional to its fitness and aging timescale. Although we found sub-linear preferential attachment in the NYT data, Figure~5B in the main text shows that \eref{average_impact_dynamics} mathces the impact dynamics in the NYT data well despite being derived assuming that preferential attachment is absent.

The generalized form of \eref{model_continuous} is
\begin{equation}
\label{model_continuous_PA}
\frac{\dd \overline{c_i(t)}}{\dd t} = \eta_i F[c_i(t)]\exp[-(t-t_i)/\varTheta_i]
\end{equation}
where $F[c_i(t)]$ is a preferential attachment term. Standard preferential attachment is represented with $F[c_i(t)] = C + c_i(t)$ where $C$ is an additive term which is necessary as all articles have initially $c_i(t_i)=0$. Non-linear preferential attachment is represented with, for example, $F[c_i(t)] = C + c_i(t)^{\beta}$. Preferential attachment with saturation is represented with a piece-wise function of $c_i(t)$. The resulting distribution of article impact for various choices of $F[c_i(t)]$ is studied in Section~\ref{sec:PA} where we show that when the preferential attachment dependence in $F[c_i(t)]$ has a cut-off, the tail of the impact distribution has the same distribution as $\eta_i\varTheta_i$. This contrasts with the results presented in~\cite{medo2011temporal} where the authors show that exponentially distributed fitness combined with preferential attachment gives rise to a power-law degree distribution.

\clearpage
\section{Estimating individual aging timescales}
\begin{figure}[h!]
\centering
\includegraphics[scale=0.6]{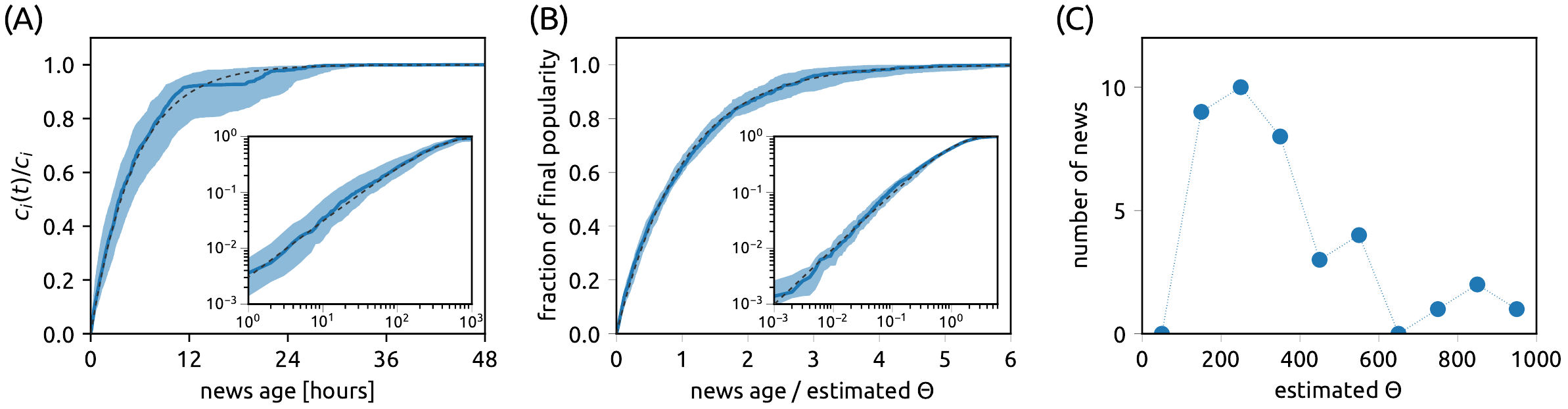}
\caption{\textbf{Commenting dynamics in the BBC data.} (A) The dynamics of article impact in the real time. The solid line shows the median fraction of the final popularity at given article age; the shaded region shows the 20th--80th percentile range of the observed popularity fraction values. The dashed line corresponds to \eref{average_impact_dynamics} with the aging timescale $\varTheta=305\,\text{min}$. (B) As panel (A) but time is rescaled with the individual aging timescale $\varTheta$ for each article (same Figure~5A in the main text). The inset focuses on the early time after the article appearance using a log-log scale. (C) The distribution of the individual aging timescales among the articles. As in Figure~5 in the main text, we include morning hit articles in the analysis here.}
\label{fig:SI_indiv_timescales_BBC}
\end{figure}

\begin{figure}[h!]
\centering
\includegraphics[scale=0.6]{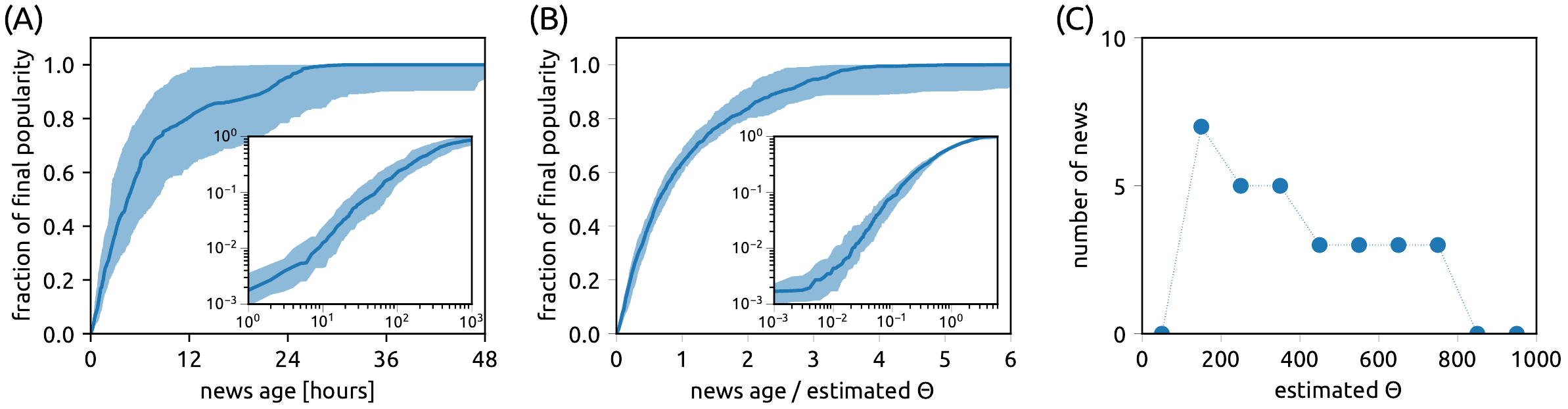}
\caption{\textbf{Commenting dynamics in the NYT data.} All panels as in Figure~\ref{fig:SI_indiv_timescales_BBC}.}
\label{fig:SI_indiv_timescales_NYT}
\end{figure}

Note that we assume an individual aging timescale, $\varTheta_i$, for each article. One can ask whether this generalization is really necessary, if universal aging timescale $\varTheta$ would not yield a similar level of agreement between the model and the empirical data. We illustrate this in Figures~\ref{fig:SI_indiv_timescales_BBC} and \ref{fig:SI_indiv_timescales_NYT} below where the left panels use real time on the $x$ axes and the middle panels use rescaled time $(t - t_i) / \varTheta_i$ on the $x$ axes. The aging timescales are estimated by minimizing the Kolmogorov-Smirnov statistic between the real course of $c_i(t)$ and the theoretically expected curve given by \eref{average_impact_dynamics}. For the BBC data, the lowest mean KS is achieved with the universal timescale of 300 minutes which agrees with the timescale obtained by fitting the aging curve in Figure~4A in the main text; the mean KS static is then 0.18 and 9 articles (out of the 39 used to obtain the figure) have the KS statistic below 0.1. With individually estimated timescales, the mean Kolmogorov-Smirnov statistic reduces by the factor of two to 0.08 and 34 articles have the KS statistic below 0.1. For the NYT data, the lowest mean KS is achieved with the universal timescale of 330 minutes; the mean KS is then 0.27 and 5 article discussions out of 46 have the KS statistic below 0.1 (the values are 0.29 and 4 for the timescale 230 obtained by fitting the aging curve in Figure~4B in the main text). With individually estimated timescales, the mean KS reduces to 0.12 and 30 articles have the KS statistic below 0.1. The improvement achieved by using \eref{model_continuous_PA} which includes preferential attachment is minor both in terms of the mean KS as well as the number of articles with KS below 0.1. To summarize, the commenting dynamics described by \eref{average_impact_dynamics} fits the empirical data well even when some limited effects of preferential attachment can be observed in the NYT data (as shown in Figure~3B in the main text).

\clearpage
\section{Interplay between exponential aging and circadian user activity patterns for the NYT data}
\begin{figure}[h!]
\centering
\includegraphics[scale=0.65]{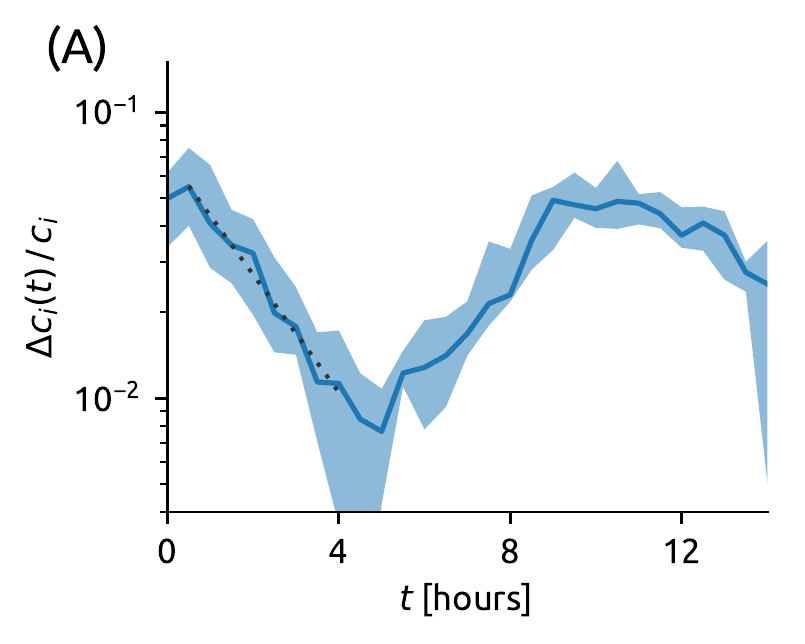}
\includegraphics[scale=0.65]{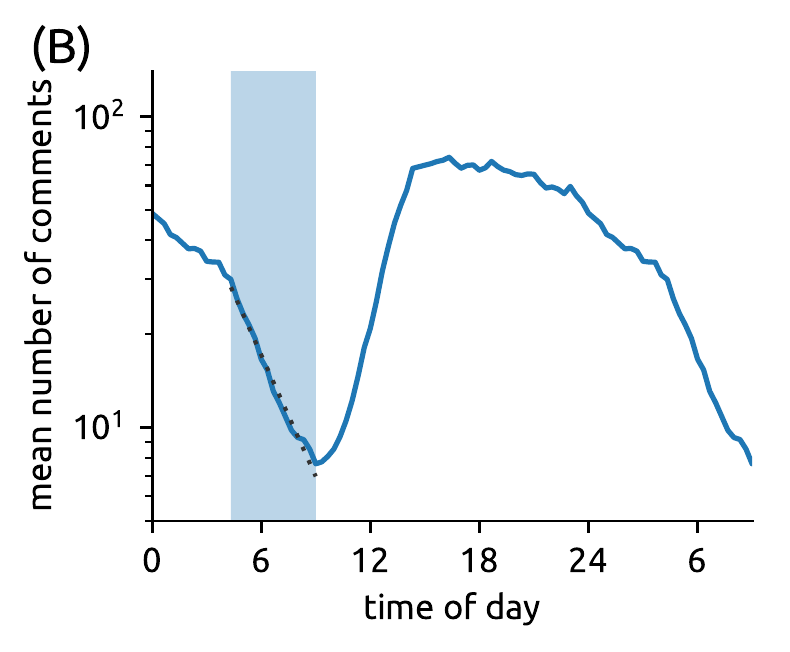}
\caption{\textbf{Interplay between exponential aging and circadian user activity patterns in the NYT data.} (A) The number of new comments of an article, $\Delta c_i(t,\Delta t)$, normalized by the final number of comments, $c_i$, as a function of its age, $t$, for evening hit articles (published between 3 am and 6 am GMT). The dotted line indicates the linear fit for age 30--240 minutes (fitted timescale 126 minutes). The age bin size here is 30 minutes to achieve better statistics for high article age. (B) The variation of the mean number of comments in 20 minute intervals during the day in the NYT data. Fitted timescale of the exponential decay between 4 am and 9 am GMT is 200 minutes.}
\label{fig:interplay}
\end{figure}

Since the overall user activity varies in the NYT data in a similar way as it does in the BBC data, its impact on article impact dynamics can be studied analogously to the section ``Interplay between exponential aging and circadian user activity patterns'' in the main text where results for the BBC data are presented. While the aging timescale of morning articles in the NYT data is $\Theta_A=230\,\text{min}$ (see Figure~4B in the main text), Figure~\ref{fig:interplay}A shows that the aging timescale is significantly shorter, 126 minutes, for evening articles. Figure~\ref{fig:interplay}B shows that the overall user activity decays in the with the fitted timescale of $\Theta_U=200\,\text{min}$. Using Eq.\,(5) from the main text, the corresponding joint timescale is
$$
1/\Theta_J = 1/(230\,\text{min}) + 1/(200\,\text{min}) \quad\implies\quad \Theta_J=107\,\text{min}
$$
which is close to the actual fitted aging time scale of evening articles, 126 minutes.

Note that the NYT data are less suitable for this analysis than the BBC data for two main reasons. Firstly, evening user activity decay is piece-wise with the first slower part between 11 pm and 4 am and the second faster part between 4 am and 9 am (we use GMT). This is probably due to the large geographical size of USA which mainly covers four time zones (from the east coast to the west coast). In our analysis, we use the more pronounced second part when, however, the overall user activity is low which makes our statistical estimates more noisy. Secondly, as many articles appear on BBC Sport in the evening, Figure~6B in the main text is built on data from 56 evening hit articles. By contrast, the analogous Figure~\ref{fig:interplay}A uses only 13 NYT hit articles that appear in the evening which makes statistical estimates less precise. Thirdly, daily variations are greater in the BBC data (as measured by the ratio between the maximal and minimal user activity during the day, for example) which makes this dataset more suitable for analyzing the impact of daily variations on the dynamics of article impact.

\end{document}